\documentclass[journal=jpclcd, manuscript=letter, layout=twocolumn]{achemso}

\usepackage[version=3]{mhchem} 

\usepackage{tikz}
\usetikzlibrary{shapes,arrows,graphs}
\usepackage{amsmath}
\usepackage{amssymb}
\usepackage{physics}

\def \be{\begin{equation}}
\def \ee{\end{equation}}
\def \ba{\begin{align}}
\def \ea{\end{align}}

\author{Ayush Saurabh}
\email{asaurabh@asu.edu}
\affiliation[CBP]
{Center for Biological Physics, Department of Physics, Arizona State University, Tempe, AZ 85287, USA}

\author{Stefan Niekamp}
\email{niekamp.stefan@googlemail.com}
\affiliation[MGH]
{Massachusetts General Hospital, Boston, MA 02114 (Current Address)}
\alsoaffiliation[UCSF]
{Department of Cellular and Molecular Pharmacology, University of California, San Francisco,
CA 94158}

\author{Ioannis Sgouralis}
\email{isgoural@utk.edu}
\affiliation[UTK]
{Department of Mathematics, University of Tennessee, Knoxville, TN 37996, USA}

\author{Steve Press\'e}
\email{spresse@asu.edu}
\affiliation[CBP]
{Center for Biological Physics, Department of Physics, Arizona State University, Tempe, AZ 85287, USA}
\alsoaffiliation[SMS]{School of Molecular Sciences, Arizona State University, Tempe, AZ 85287, USA}

\title[An \textsf{achemso} demo]
  {Modeling Non-additive Effects in Neighboring Chemically Identical Fluorophores
  } 

\abbreviations{ROI}
\begin{tocentry}
\includegraphics[width=1.0\textwidth]{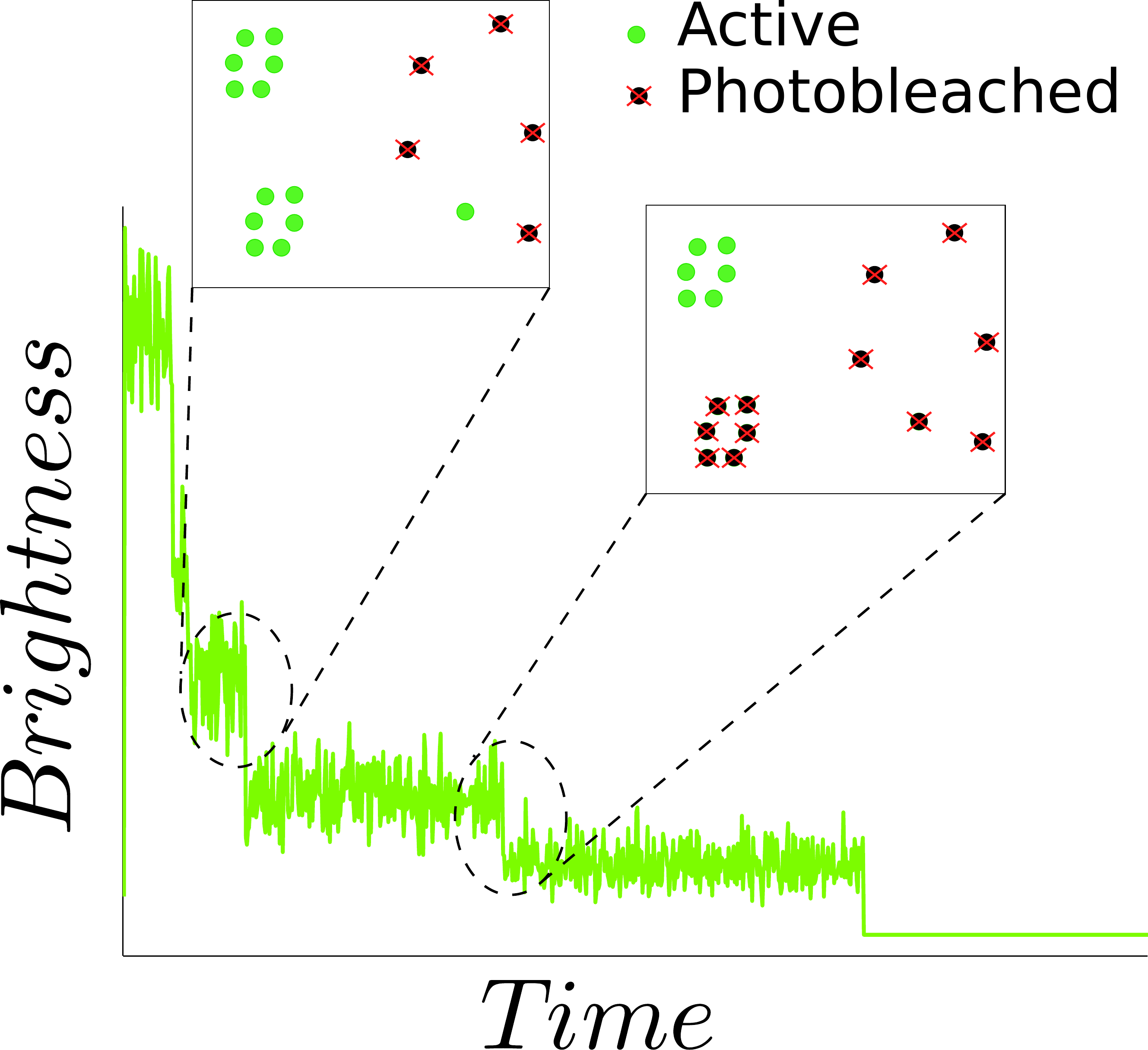}
\end{tocentry}

\begin{document}
\begin{abstract}
	Quantitative fluorescence analysis is often used to derive chemical properties, including stoichiometries, of biomolecular complexes. One fundamental underlying assumption in the analysis of fluorescence data---whether it be the determination of protein complex stoichiometry by super-resolution, or step-counting by photobleaching, or the determination of RNA counts in diffraction-limited spots in RNA fluorescence {\it in situ} hybridization (RNA-FISH) experiments ---is that 
	fluorophores behave identically and do not interact. However, recent experiments on fluorophore-labeled DNA-origami structures such as fluorocubes have shed light on the nature of the interactions between identical fluorophores as these are brought closer together,
	thereby raising questions on the validity of the modeling assumption that fluorophores do not interact. Here, we analyze photon arrival data under pulsed illumination from fluorocubes where distances between dyes range from 2-10 nm. We discuss the implications of non-additivity of brightness on quantitative fluorescence analysis.  \end{abstract}


\textit{Introduction.} Fluorescent labels have been critical in allowing us to discriminate between homogeneous background and biomolecules of interest \cite{10.1042/BSR20170031}. For instance, they have been useful in determining the locations of molecules within neighboring regions by exploiting the nonlinear response (fluorophore activation) 
of fluorophores to incoming light \cite{doi:10.1126/science.1127344, Hell:94}. They have also allowed us to determine the stoichiometry of protein complexes based on a spot's emission intensity \cite{Tsekouras2016, Bryan2020.09.28.317057, Arant2014, Ulbrich2007}. In particular, we note that photobleaching event counting experiments, where 
fluorophores are stochastically deactivated, lead to step-like patterns in  brightness traces \cite{doi:10.1091/mbc.e14-06-1146}. 

Quantitative fluorescence analysis experiments have been used extensively throughout biophysics to quantify the 
stoichiometry of a number of complexes involved, for example, in bacterial  flagellar  
switch \cite{Delalez11347}, eukaryotic  flagella \cite{10.1083/jcb.200812084}, point centromere \cite{10.1083/jcb.201106036, 10.1083/jcb.201106078}, mammalian  neurotransmitter  receptors 
\cite{McGuire2012},  human  calcium  channels \cite{Demuro17832}, transmembrane 
$\alpha$-amino-3-hydroxy-5-methyl-4-isoxazolepropionic acid receptor-regulatory 
proteins \cite{Hastie5163},  T4  bacteriophage  helicase  loader  protein 
\cite{Arumugam2009},  bacterial  oxidative  phosphorylation  com-plexes 
\cite{LLORENTEGARCIA2014811}, microRNAs in processing bodies 
\cite{PITCHIAYA2013188, pitchiaya2014}, RNAs in a bacteriophage DNA-packaging 
motor \cite{shu2007}, and other membrane proteins and protein complexes \cite{Leake2006, Das2007}.

However, a quantitative analysis of the stoichiometry of a protein complex \cite{Ulbrich2007, Singh2020} or the enumeration of the number of fluorophores within a diffraction-limited spot \cite{Bryan2020.09.28.317057}, or other chemical properties of a system tagged with identical fluorophore labels \cite{Dhara2020}, unavoidably requires simplifying assumptions. One critical assumption pervading such analyses is that chemically identical fluorophores do not interact and therefore are photo-emissively identical as well. Here, we 
reassess this fundamental assumption at the basis of data derived from  
modern techniques such as PALM \cite{doi:10.1126/science.1127344, Sengupta2012, 
Lee17436}, STORM \cite{STORM}, photobleaching event counting 
\cite{doi:10.1091/mbc.e14-06-1146}, RNA-FISH \cite{Amann2008}, and others \cite{Gru_mayer_2019, 
DANKOVICH2021102134}. 

To carefully isolate and analyze the non-additive effects of fluorophore interactions, we must account for a number of factors including noise generated by acquisition devices (cameras or single photon detectors) \cite{Mandracchia2020, emCCD2013}, the quantized nature of photon emission (shot noise), and background emissions. Therefore, an accurate assessment of non-additive effects as well as a quantitative treament of fluorescence data requires a hierarchical mathematical treatment of the stochastic effects arising from the contributions mentioned above.  

Previous studies have quantified some of these contributions. For example, 
instrument response function (IRF) \cite{doi:10.1063/1.3095677, 
SZABELSKI2009153, lakowicz2013principles}, shot noise 
\cite{Bryan2020.09.28.317057, doi:10.1021/acs.jpcb.8b09752}, and noise from 
background  fluorescence \cite{Coffman2012, Ulbrich2007}. However, 
incorporating the effects of unintended interactions among identical 
fluorophores and biomolecules \emph{remains an open challenge.}  As a result, a 
forward model is required in order to develop an inverse strategy to infer 
protein complex stoichiometry or even the enumeration of multiply-labeled RNA, 
especially when these RNA are in closely-spaced physical regions, from RNA-FISH 
experiments \cite{Xie2018} is yet to be proposed. In particular, the primary 
goal of this paper is not to elucidate the quantum mechanical origin of 
interactions among fluorophores but to devise a strategy to analyze imaging 
data.

Recent studies on fluorescently-labeled DNA-origami structures like fluorocubes 
\cite{Niekamp716787, schroder_scheible_steiner_vogelsang_tinnefeld_2019, 
Helmerich2022.02.08.479592} have shown to what degree some identical 
fluorophores interact when separated by distances ranging from 2-10 nm, leading 
to improved photostability \cite{Niekamp716787} but also increased 
photoblinking whose statistical signature has very recently been proposed as a 
tool to determine molecular distances below 10 nm. These experiments provide 
building blocks that may serve in constructing models of interacting 
fluorophores \cite{Helmerich2022.02.08.479592}. 

\begin{figure}[t!]
\centering
\begin{tikzpicture}
        \draw[gray,opacity=0.4,domain=0.1:pi,samples = 50] plot
        ({0.2*sin(500*\x)}, \x);
        \draw[gray,opacity=0.4,domain=0.1:pi,samples = 50] plot
        ({-0.2*sin(500*\x)}, \x);
        \draw[gray,opacity=0.4,domain=0.1:pi,samples = 50] plot
        ({0.2*sin(500*\x)+2}, \x);
        \draw[gray,opacity=0.4,domain=0.1:pi,samples = 50] plot
        ({-0.2*sin(500*\x)+2}, \x);
        \draw[gray,opacity=0.4,domain=0.1:pi,samples = 50] plot
        ({0.2*sin(500*\x)+3}, \x+1);
        \draw[gray,opacity=0.4,domain=0.1:pi,samples = 50] plot
        ({-0.2*sin(500*\x)+3}, \x+1);
        \draw[gray,opacity=0.4,domain=0.1:pi,samples = 50] plot
        ({0.2*sin(500*\x)+1}, \x+1);
        \draw[gray,opacity=0.4,domain=0.1:pi,samples = 50] plot
        ({-0.2*sin(500*\x)+1}, \x+1);
        \fill[red,opacity = 0.6] (0,0) ellipse (0.8cm and 0.4cm) node[black]
        {$F_1$};
        \fill[red,opacity = 0.6] (0,3) ellipse (0.8cm and 0.4cm) node[black]
        {$F_6$};
        \fill[red,opacity = 0.6] (2,0) ellipse (0.8cm and 0.4cm) node[black]
        {$F_2$};
        \fill[cyan,opacity = 0.6] (2,3) ellipse (0.8cm and 0.4cm);
        \fill[red,opacity = 0.6] (3,1) ellipse (0.8cm and 0.4cm) node[black]
        {$F_3$};
        \fill[red,opacity = 0.6] (3,4) ellipse (0.8cm and 0.4cm) node[black]
        {$F_4$};
        \fill[red,opacity = 0.6] (1,4) ellipse (0.8cm and 0.4cm) node[black]
        {$F_5$};
        \draw[|-|,thick] (0,-1) -- (2,-1) node[pos=0.5,above] {a = 2-4 nm};
        \draw[|-|,thick] (3.5,-0.2) -- (4.5,0.8) node[pos=0.5,sloped,above] {2-4
        nm};
        \draw[|-|,thick] (-1.2,0.0) -- (-1.2,3.0) node[pos=0.5,sloped,above] {b
        = 5.4  nm};
\end{tikzpicture}

        \caption{\textbf{Fluorocube.} In a fluorocube, six fluorescent dyes (red) are attached to the ends of the four 16
        bp DNA helices. One position is usually reserved for a functional tag (blue) to be linked to a molecule of interest \cite{Niekamp716787} and last remaining corner is left unlabeled.}
        \label{fluorocubeDiagram}
\end{figure}
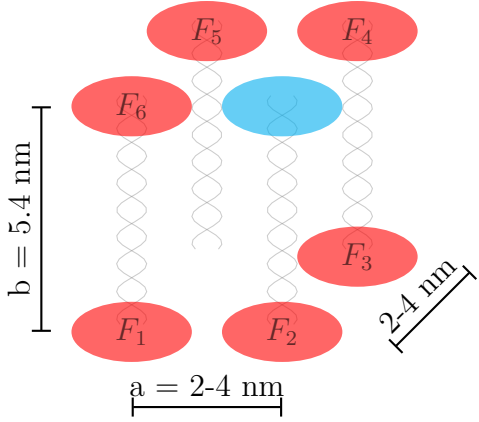

Fluorocubes \cite{Niekamp716787} are constructed using four 16 
base-pair (bp) long double-stranded DNA (dsDNA) as a highly photostable option 
for collecting long trajectories of labeled biomolecules.  The four DNA helices are assembled together in such a way that the fluorophores attached to the ends of the helices form the corners of a cuboid (see Fig.~\ref{fluorocubeDiagram}) separated by 2-6 nm. One of the corner sites is usually kept reserved for a tag that binds to proteins at specific locations and the last remaining one is left unlabeled.

The photophysical properties of these fluorocubes were studied and contrasted with cubes labeled with only one dye (single-dye cubes) and with six dyes that are separated by larger distances of 6-10 nm (large cubes) as well \cite{Niekamp716787}. It was found that these 
properties significantly vary among fluorocubes depending on the species of the dyes used. Overall, most fluorocube configurations blinked less and exhibited reduced photobleaching rates compared to their single-dye counterparts. Even the DNA scaffolding itself 
resulted in improved photostability as single-dye cubes emitted
many fold more photons in many cases with longer lifetimes by comparison to a single dsDNA helix coupled to one dye \cite{Niekamp716787}. Concretely, increasing the dye separations in such cubes from $\sim$
2-6 nm to $\sim$ 6-10 nm showed that photobleaching time decreases with size, indicating that dye-dye interactions play an important role in the fluorocube photophysics. 

Here, we make an attempt at building a model of fluorophore interactions in order to derive quantities such as transition rates between states and excitation probabilities from single photon arrivals. These quantities are then estimated for the lowest energy states of fluorocubes from experimental data. We also consider cubical structures of different sizes (single-dye cubes and large cubes, respectively) to demonstrate the changes in photophysics resulting from changes in interactions between dyes. We finally illustrate the consequences of these interactions by analyzing synthetically generated photobleaching event counting traces.

\textit{State Space.} We first develop a quantum mechanical model to describe and label all possible non-degenerate states of a fluorocube. In an effort to simplify our modeling, we make the following physical assumptions based on typical time scales and data collection procedure (see experimental methods section):
\begin{enumerate}
	\item the excitation pulse is infinitesimally narrow ($\sim 12$ ps 
		wide) compared to other time scales of the problem,

	\item the probability of simultaneously exciting two or more dyes
		by a single pulse is zero since events where two photons are emitted 
		are rarely observed ($\sim$ 1 per $10^4$ photon events or less),

	\item the duration between pulses is long enough ($\sim 80$ ns) for the 
		fluorocubes to return to the ground state before the next pulse 
		($<4$ ns average lifetime),

	\item fluorophores 
	are well approximated by a two state system with a ground and an excited
                state,
                
     \item photoblinking effects, induced by visiting rarely occupied states such as triplet states, are ignored for simplicity as we are primarily interested in interactions among fluorophores in their ground and first excited states \cite{ref1}, and
     
     \item distances between the dyes of a fluorocube do not vary during the experiments, and the upper and lower faces of a fluorocube as seen in Fig.~\ref{fluorocubeDiagram} are squares with equal separation between adjacent dyes. This may not always be true as the distances between the dyes are affected by the surrounding chemical environment \cite{Niekamp716787} and cannot be accurately determined dynamically with currently available techniques.
\end{enumerate}

The quantum mechanical nature of interactions between fluorophores is well 
known in the case of F\"orster resonance energy transfer (FRET) 
\cite{10.3389/fphy.2019.00100}. For this reason, we use a similar approach to 
develop a model for fluorocubes and postulate the following Hamiltonian
        \be
        H(t) = \sum_{i = 1}^6 H_i + H_{int} + \delta H(t),
        \label{hamiltonian}
        \ee
where $H_i$ is the Hamiltonian for individiual dyes,   $H_{int}$ is the 
Hamiltonian for interactions among dyes, and $\delta H(t)$ is the time dependent 
perturbation Hamiltonian that models excitation with light, respectively. This Hamiltonian is symmetric under a transformation swapping the dye labels, $(F_1,F_2,F_3) \rightarrow (F_6,F_5,F_4)$, as 
seen in Fig.~\ref{fluorocubeDiagram}. This symmetry can be used to group all 
the energy eigenstates of the fluorocube into a collection of non-degenerate 
energy eigenstates, without the need to identify all specific interactions 
entering the full Hamiltonian for a complex multi-molecular system.  

In total, we have $2^6 = 64$  energy eigenstates for a collection of six 
two-state systems. For a \emph{large cube} with distances between dyes varying from 6 to 10 nm, $H_{int} \approx 0$. Therefore, in this case alone, all fluorophores act effectively independently leading to maximum allowable degeneracy among the system's energy eigenstates. 

The 64 states of such a large cube can be grouped according 
to the number of dyes excited at any instant: 6 degenerate states for one and 
five-dye excitations each, 15 degenerate states for two and four-dye
excitations each, 20 degenerate states for three-dye excitations, and one state 
for ground and six-dye excitations each, respectively. 

Interactions among fluorophores, however, reduce this degeneracy. Such splitting among the degenerate states of the system is akin to level splitting in the Zeeman effect \cite{griffiths_schroeter_2018} and is fully dictated by the 
symmetries of the interaction Hamiltonian. Here, we focus on the geometrical symmetries of a \textit{fluorocube} where interactions are expected to be significant and do not consider fine-splitting induced by the asymmetries of the individual molecules in the assembly.

From the geometrical symmetries of a fluorocube (see Fig.~\ref{fluorocubeDiagram}), it can be seen that the degenerates states will split into multiple levels; 6 degenerate one-dye excitation states split 
into 3 levels, 15 degenerate two-dye excitation states split into 9 levels,  20 
degenerate three-dye excitation states split into 10 levels, 15 degenerate 
four-dye excitation states split into 9 levels, and 6 degenerate five-dye 
excitation states split into 3 levels, respectively. 

We can mathematically represent these 35 non-degenerate excitations as 
\begin{align}
	\boldsymbol{\sigma} = \{
	&\ket{1}^{(1)},\ket{1}^{(2)},\ket{1}^{(3)}, \nonumber \\
	&\ket{2}^{(1)},\ket{2}^{(2)},\ket{2}^{(3)},\ldots,\ket{2}^{(9)}, \nonumber \\ 
	&\ket{3}^{(1)},\ket{3}^{(2)},\ket{3}^{(3)},\ldots, \ket{3}^{(10)}, \nonumber \\ 
	&\ket{4}^{(1)},\ket{4}^{(2)},\ket{4}^{(3)},\ldots,\ket{4}^{(9)}, \nonumber \\
	&\ket{5}^{(1)},\ket{5}^{(2)},\ket{5}^{(3)}, \nonumber \\
	&\ket{6}^{(1)} \},
	\label{setofstates}
\end{align}
where the labels inside the bra-ket notation indicates the number of dyes 
excited by the pulses and the superscripts only run over the non-degenerate 
states. This labeling simplification can be done because degenerate states 
cannot be distinguished from each other on account of similar lifetimes. We 
keep the ground state $\ket{0}$ separate from this set for convenience as it 
doesn't require a stochastic treatment, as discussed later.

In an experiment, a sample containing a collection of fluorescent probes of the same type is illuminated by a total of $N$ laser pulses. We label the inter-pulse windows between times $t_{n-1}$ and $t_{n-1}+\Delta$ with $n$, where $\Delta$ is the window size. Since we assume that fluorocubes are always in the ground state at the beginning of these inter-pulse windows, the photon arrival times can be considered independent and identically distributed (iid). We represent these measurements with $\mu_n$.  

Now that we have defined our state space and measurements, we postulate a forward model to generate these measurements and then design an inverse strategy to learn the parameters of interest. Similar mathematical formulations for fluoresence data have been recently developed and applied extensively for the modeling and analysis of experimental findings \cite{SGOURALIS20172117, SGOURALIS20172021, Jazani2019, doi:10.1063/1.5008842, KILIC2021409, PhysRevX.10.011021, KILIC2021100409}.

\textit{Forward Model.} A detected photon may have been emitted by some 
background source or by a fluorescent probe in the sample. Therefore, in 
addition to the quantum states for the fluorescent probes, we introduce a 
background state $\ket{b}$ that allows for a simple mathematical treatment of 
the background emissions. The photons emitted by such a state are uniformly 
distributed over the inter-pulse window with a probability of emission $\pi_b$.
This uniform assumption is validated
from experimental data under the assumption that most photons whose arrivals 
exceed 40 ns are background emissions; see Fig.~\ref{distributions} (b) with 
almost equal photon counts per bin beyond 40 ns.


If the fluorescent probes get excited, the system's states before photons are emitted are represented by $s_n$ which take values from the set of states in Eq.~\ref{setofstates}. Given the 
infinitesimal width of the excitation laser, the probabilities of exciting the 
fluorocubes to states $\boldsymbol{\sigma}$ can be collected into a constant 
probability vector $\boldsymbol{\pi_\sigma}$. These probabilities are labeled 
$\pi_{i}$ (where $i = 1,\ldots, 35$ in order of increasing excitation level) 
corresponding to the excited states in Eq.~\ref{setofstates}. Additionally, the 
probability of staying in the ground state is labeled as $\pi_0$, and since it 
can be computed directly from the number of inter-pulse windows without photon 
detections, it is not treated as a random variable. 

Following excitation, the system's state is randomly chosen from a Categorical distribution (a generalization of the Bernoulli distribution for many options)
\begin{equation}
 s_n |\boldsymbol{\pi_{\sigma}}, \pi_b \sim \mathbf{Categorical}_{\boldsymbol{\sigma}, \ket{b}}( \boldsymbol{{\pi}_{\sigma}}, \, \pi_b).
\end{equation}

On the other hand, inter-pulse windows can either be empty with no photons, or have photon arrival times that are either exponentially or uniformly distributed (for background) continuous random variables
\begin{align}
\mu_n & | s_n, \boldsymbol{\lambda_{\sigma}} \sim \nonumber \\ & \begin{cases}
\emptyset, & s_n = \ket{0} \\
\mathbf{Exponential}( \lambda_{s_n}), & s_n \ne \ket{0}, \, s_n \ne \ket{b} \\
\mathbf{Uniform}(0, \Delta), & s_n = \ket{b}, 
\end{cases}
\end{align}
where $\lambda_{s_n}$ is the escape rate for the corresponding excited state $s_n$. All escape rates corresponding to the 35 non-degenerate excited states of Eq.~\ref{setofstates} are labeled $\lambda_{i}$ for the states in Eq.\ref{setofstates} and collected in the set $\boldsymbol{\lambda_\sigma}$. 

\begin{figure*}[h!]
\includegraphics[width=0.8\textwidth]{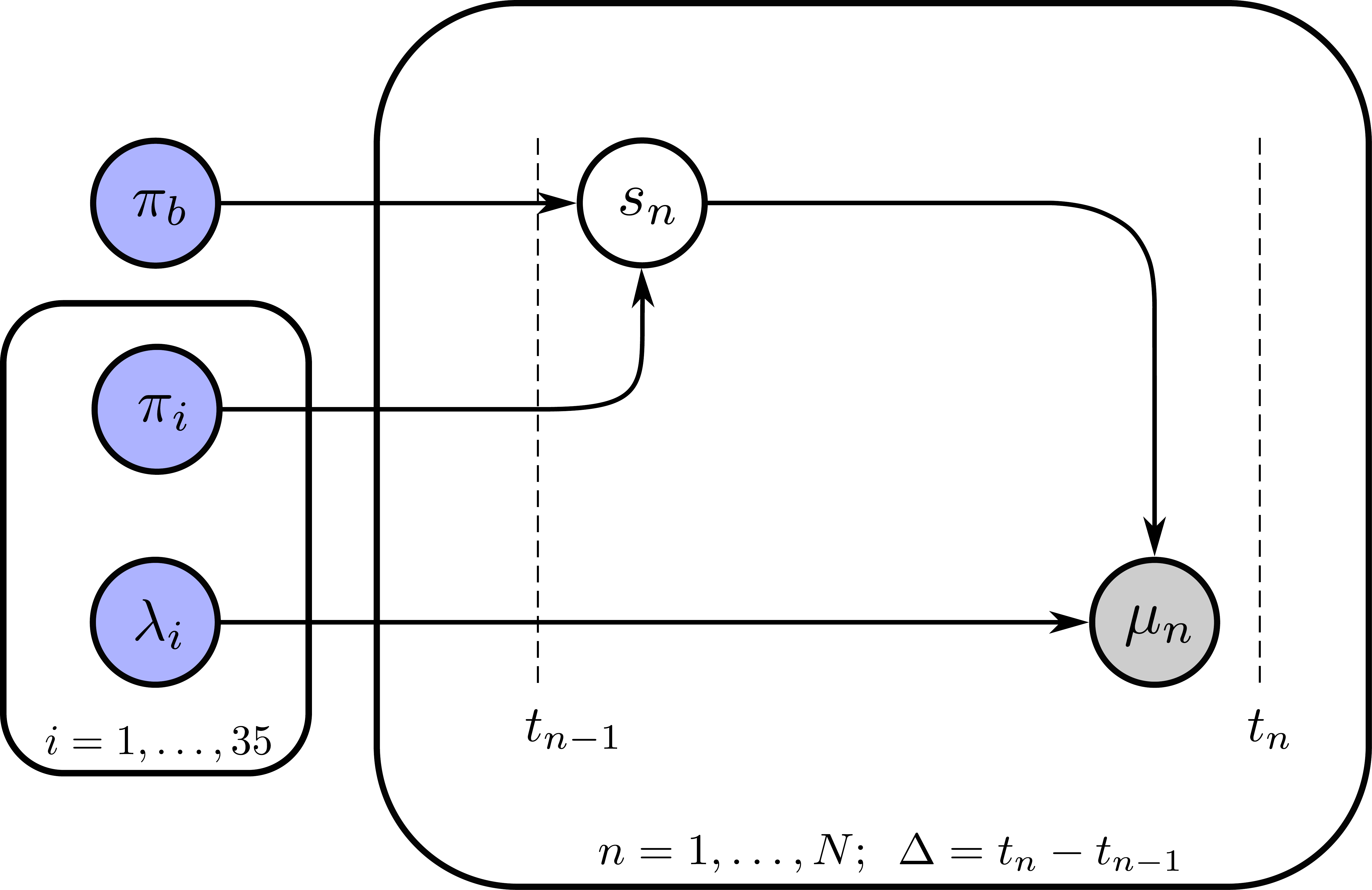}
	\caption{ \textbf{Graphical Model.} A graphical model depicting the 
	random variables and parameters involved in the generation of photon 
	arrival data for fluorocubes. Circles represent the random variables 
	involved in the inverse model.  Circles shaded in blue correspond to 
	the parameters of interest (rates and probabilities), and the one 
	shaded in gray corresponds to the measurements, while the unshaded 
	circles represent the hidden/latent variables of the model.  The arrows 
	represent conditional dependence among variables. The two plates 
	enclosing the excitation rates/probabilities and hidden 
	variables/measurements indicate that these variables are repeated over 
	indices $i$ and $n$, respectively.
	}
	\label{graphicalmodel}
\end{figure*}

The forward model described above can be visualized using the graphical 
model shown in Fig.~\ref{graphicalmodel} \cite{bishop}. Based on this formulation, the likelihood function for the arrival time of a photon is given by a mixture of probability distributions
\begin{align}
        p(\mu_n | \boldsymbol{\lambda_\sigma}, & \pi_0, \boldsymbol{{\pi}_{\sigma}}, \pi_b)   = \pi_{0}\delta_{\emptyset}(\mu_n) \nonumber \\
	&+  \pi_{1}\,\mathbf{Exponential}(\mu_n; \lambda_1)  \nonumber \\
        &+ \ldots \nonumber \\
&+ \pi_{{35}}\,\mathbf{Exponential}(\mu_n; \lambda_{35}) \nonumber \\ &+ \pi_{{b}}\,\mathbf{Uniform}(\mu_n; 0, \Delta)
\label{likelihood}
\end{align}
and, for a set of $N$ pulses, the likelihood becomes a product of many such mixtures
\begin{equation}
    p(\mu_{1:N} | \boldsymbol{\lambda_\sigma}, \pi_0, \boldsymbol{\pi_{\sigma}}, \pi_b) = \prod_{n=1}^N  p(\mu_n | \boldsymbol{\lambda_\sigma}, \pi_0,\boldsymbol{\pi_{\sigma}}, \pi_b).
    \label{posterior}
\end{equation}

    \textit{Inverse Model.} To learn these rates and excitation probabilities from the photon arrival data, we use Bayesian inference. Using Bayes' theorem, the probability distribution over these parameters \cite{bishop, sivia, van_de_schoot_bayesian_2021} (termed the posterior)
is given by 
\begin{align} p(\boldsymbol{\lambda_\sigma},\boldsymbol{\pi_{\sigma}}, \pi_b| \mu_{1:N} , \pi_0  ) &\propto p(\mu_{1:N}
	| \boldsymbol{\lambda_\sigma}, \pi_0 ,\boldsymbol{\pi_{\sigma}}, \pi_b) \nonumber \\
	&\times p(\boldsymbol{\lambda_\sigma})  p(\boldsymbol{\pi_{\sigma}}, \pi_b).
	\label{posterior}
\end{align}
 Those distributions multiplying the likelihood, 
 $p(\boldsymbol{\lambda_\sigma})$ and  $p(\boldsymbol{\pi_{\sigma}}, \pi_b)$ 
 are priors and selected on the basis of computational convenience and in whose 
 domains the associated parameters exist \cite{sivia}. Additionally, the 
 influence of these priors on the posterior distribution is minimal when large 
 amounts of data, incorporated through the likelihood, are available. 

We use the following priors for the transition rates and excitation 
probabilities
\begin{align}
        \lambda_{i} & \sim \mathbf{Gamma} (A, \lambda_{ref}/ A), \text{ and}  \\ 
	\boldsymbol{\pi_{\sigma}}, \pi_b& \sim 
 	\mathbf{Dirichlet}_{\boldsymbol{
	\sigma}, \, \ket{b}} ( \boldsymbol{\alpha}_{\boldsymbol{\sigma}, b}),
 \end{align}
 where we choose $A = 1$ and $\lambda_{ref} = 1$ ns$^{-1}$, and set all elements of  the vector $ \boldsymbol{\alpha}_{\boldsymbol{
	\sigma}, b}$ to be 1. 

To sample transition rates and excitation probabilities from the 
posterior distribution above, we use the Gibbs algorithm \cite{bishop}, which involves sampling individual parameters from their respective conditional posteriors
\begin{align}
    \lambda_i &\sim p(\lambda_i|\boldsymbol{\lambda_\sigma} \backslash \lambda_i , \pi_0, \boldsymbol{\pi_{\sigma}}, \pi_b , \mu_{1:N} ), \text { and}  \label{conditionals1}\\
    & \boldsymbol{\pi_{\sigma}}, \pi_b \sim p(\boldsymbol{\pi_{\sigma}}, \pi_b |\boldsymbol{\lambda_\sigma}, \pi_0, \mu_{1:N} ), \label{conditionals2}
\end{align}
where the parameters following a backslash ``$\backslash$" are excluded from the preceding set. The above Gibbs sampling scheme  
still requires brute-force Metropolis-Hastings algorithm for each of the three steps as none of the associated conditional distributions allow for direct sampling. We illustrate results from the above sampling scheme in supplementary section 2.

The computational cost of brute force sampling does now help motivate the (more complex)  sampling scheme now presented below. 

To generate conditional posteriors in a Gibbs sampling scheme from which we can sample directly, we now exploit the conjugacy of Exponential-Gamma and Multinomial-Dirichlet likelihood-prior pairs. To exploit these properties, we must
de-marginalize the distribution in Eq.~\ref{posterior} over the system's states $s_n$ as follows
\begin{align} 
p(\boldsymbol{\lambda_\sigma},& \boldsymbol{\pi_{\sigma}} ,\pi_b | \mu_{1:N}, \pi_0 ) 
	\propto \nonumber \\ & \left(
	\sum_{s_{1:N}} p(\mu_{1:N},s_{1:N}
	| \boldsymbol{\lambda_\sigma}, \pi_0, \boldsymbol{\pi_{\sigma}}, \pi_b) \nonumber
	\right)
	\\
	& \hspace{24mm} \times p(\boldsymbol{\lambda_\sigma}) \, p(\boldsymbol{\pi_{\sigma}}, \pi_b).
\end{align}
This de-marginalization creates a new set of variables, $s_{1:N}$, which we must now also sample by supplementing the Gibbs algorithm presented from Eqs~\ref{conditionals1}-\ref{conditionals2}. 

On account of these conjugacy conditions, the updated Gibbs sampling algorithm is now as follows
\begin{align}
s_n  & \sim p(s_n \, | \,\boldsymbol{\lambda_\sigma}, \pi_0 ,\boldsymbol{\pi_{\sigma}}, \pi_b, s_{1:N} \backslash s_n  , \mu_{1:N} ) \nonumber  \\
& \propto \mathbf{Categorical}_{\boldsymbol{\sigma}, \, \ket{b}}\Big(\mathbf{Exponential}(\mu_n| \lambda_1) \, \pi_1, \nonumber \\
& \ldots,\,\,
\mathbf{Exponential}(\mu_n| \lambda_{35}) \, \pi_{35}, \pi_b/\Delta \Big)
\end{align}
\begin{align}
    \boldsymbol{\pi_{\sigma}}, \pi_b & \sim p(\boldsymbol{\pi_{\sigma}} \, | \,\boldsymbol{\lambda_{\sigma}},\pi_0, s_{1:N},  \mu_{1:N}) \nonumber \\
    & \propto \mathbf{Dirichlet}_{  \boldsymbol{
	\sigma}, \, \ket{b}}(\,\boldsymbol{\alpha}_{\boldsymbol{
	\sigma}, b}+\boldsymbol{\eta}_{\boldsymbol{\sigma}, b}), \,\, \text{and}
\end{align}
\begin{align}
   \lambda_i & \sim p(\lambda_i \,|\,\boldsymbol{\lambda_\sigma} \backslash \lambda_i, \pi_0 ,\boldsymbol{\pi_{\sigma}}, \pi_b, s_{1:N} , \mu_{1:N} ) \nonumber \\
    & \propto  \mathbf{Gamma} \big( \, A+\sum_{n = 1}^N {\Delta_i(s_n)}, \, \lambda_{ref}/A \nonumber \\ &\hspace{30mm} + \sum_{n = 1}^N \mu_n^{\Delta_i(s_n)} \big) ,
\end{align}
where $\boldsymbol{\eta}_{\boldsymbol{\sigma}, b}$ is a vector containing the number of pulses exciting a fluorocube to a given state in $\boldsymbol{\sigma}$ and to the background state $\ket{b}$, and $\Delta_i (s_n)$ is 1 whenever $s_n = \ket{i}$ and otherwise 0, respectively.

\begin{figure*}[h!]
\includegraphics[width=1.0\textwidth]{figures/bivariate_plots_new.pdf}
	\caption{ \textbf{Experimental Data Analysis.} In row (a), we have 
	three $256 \times 256$ pixels raster-scanned images of samples 
	containing three cubes labeled with ATTO 647N dyes, each. The images on 
	the left and the right are for single-dye cubes and fluorocubes 
	illuminated with 80\% maximum laser power, respectively. However, the 
	center image for large-cubes was produced using only 30\% laser power.  
	The laser power is varied in order to obtain approximately equal photon 
	numbers per probe per  experimental run. Histograms for photon arrival 
	times (microtimes) recorded from these images are shown in row (b). In 
	row (c), each panel shows a bivariate posterior for escape rates 
	$\lambda_i$ (log scale) and corresponding probabilities $\pi_i$ as in 
	Eq.~\ref{likelihood}. We smooth out the bivariate plots using the 
	kernel density estimation (KDE) tool available in Julia. The left panel 
	shows the distribution for single-dye cubes. The distribution is 
	concentrated about $\lambda_i \approx 0.25$ ns$^{-1}$ or a lifetime of 
	around $1/\lambda_i$ $= 4.0$ ns. $\pi_b \approx 0.10$ and $\pi_0 
	\approx 0.49$ in this case. The plot for large cubes in the center 
	panel shows no change in lifetime, suggesting insignificant 
	interactions. $\pi_b \approx 0.08$ and $\pi_0 \approx 0.37$ for these 
	cubes. In the rightmost panel for fluorocubes, the peak moves towards a 
	significantly shorter lifetime of around 1.54 ns. The background 
	probability $\pi_b$ here is 0.12, slightly higher than that of 
	non-interacting cubes, and $\pi_0 \approx 0.59$.}  	
	\label{distributions}
\end{figure*}

\textit{Analysis of ATTO-647N Fluorocubes.} We now use the formulation described above to learn how photophysical properties of fluorocubes and large cubes may change as the distances between dyes are varied. The photon arrival data is acquired as a collection of these cubical fluorescent probes are illuminated by a pulsating laser around every 80 ns. The experiment is repeated many times under different intensities of the laser (see the experimental methods section).

In  Fig.~\ref{distributions}, we show captured images, microtime histograms, and three bi-variate plots of the learned distributions for transition rates (lifetimes) and excitation probabilities for three different types of cubes labeled with ATTO-647N dyes, respectively.

For single-dye cubes and large cubes --- illuminated with 80 \% and 30 \% of 
the maximum laser power, respectively, we expect the interactions to not affect 
the photophysics significantly and that is what we observe (since large cubes 
are not affected by interactions, they are expected to emit far more photons 
per probe than fluorocubes and are therefore illuminated with less power to 
have the same number of photon emissions per probe). As shown in the left and 
center panels of Fig.~\ref{distributions} (c), the most frequently sampled 
rates are of around 0.25 ns$^{-1}$  which are equivalent to lifetimes of around 
4.0 ns, close to the value specified by the manufacturer ($\approx 3.5$ ns) 
\cite{ATTO647N} and therefore used as an independent measurement here to 
compare parameters when interactions are present. This provides one form of 
validation for the formalism we are using here. Additionally, the probability 
of excitation is much higher for the large cubes compared to the single-dye 
cubes, which is expected since a large cube has 5 additional fluorophores that 
can be independently excited at the same time.

The most frequently sampled rate in the case of fluorocubes illuminated with 80 
\% laser power (rightmost panel in Fig.~\ref{distributions} (c)), where 
interactions are expected, is around 0.65 ns$^{-1}$ or an equivalent lifetime 
of 1.54 ns. This clearly indicates that photophysical properties of a 
collection of fluorophores change significantly when they are in close 
proximity. However, we do not observe any significant splitting among the 
states as we could not isolate secondary peaks. This also suggests that the 
lowest energy excited states have similar escape rates (same order of 
magnitude) and are equally likely to be excited by a pulse. To make sure that 
this is not an artifact of the computational algorithm, we demonstrate the 
algorithm's robustness via distinctly visible splitting for synthetically 
generated data for six states with precisely known lifetimes and probabilities 
of excitations (see supplementary section 1). 

It should also be noted that excitation probabilities (normalized for laser 
power and the number of fluorophores per probe) for the smaller ATTO647N 
fluorocubes are significantly lower compared to its single-dye and the larger 
six-dye counterparts. This change in excitation probability as distance between 
fluorophores changes plays a crucial role in determining the brightness 
($\pi_i/\Delta$) of the fluorescence signal. This is also confirmed in the 
experiments by \citet{Niekamp716787}(see Figs. 3 \& 8 of their supplementary 
document), where they observe that single-dye ATTO647N cubes are typically as 
bright as the six-dye ATTO647N fluorocubes or brighter if the illumination is 
strong ($\sim 1.7$ times brighter at exposure values of 2.0 $\mu$J/$\mu$m$^2$). 
This suggests that interactions significantly dampen the brightness per 
fluorophore when a collection of fluorophores comes closer together. This is 
similar to homo-FRET \cite{GAUTIER20013000} where FRET between identical 
fluorophores decreases the quantum yield (quenching) and the lifetimes of the 
individual probes. 

This dampening effect (per fluorophore) is observed to not depend on the 
species of dye used to label the cubes \cite{Niekamp716787}. To demonstrate 
this, we repeated our experiments with large cubes and fluorocubes labeled with 
Cy3 dyes. We again observe (see Fig. 3 of the supplementary) significantly 
smaller excitation probability per fluorophore for Cy3 fluorocubes compared to 
Cy3 large cubes.

\begin{figure*}[ht!]
\includegraphics[width=1.0\textwidth]{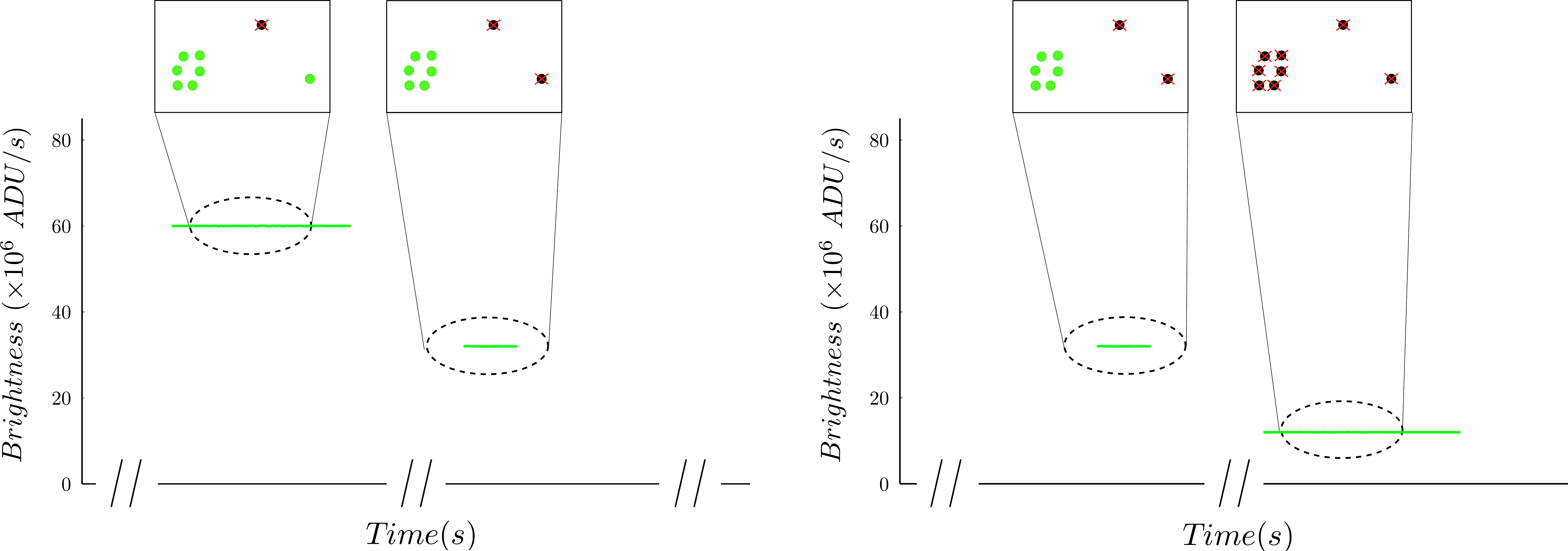}
	\caption{\textbf{Photobleaching Traces.} Here, we show two sections of 
	a synthetically generated photobleaching counting trace assuming an 
	EMCCD camera model for a collection consisting of one ATTO647N 
	fluorocube (with six tightly packed interacting dyes each) and two 
	well-separated ATTO647N (non-interacting) dyes. In (a), we see that 
	brightness is reduced when one of the non-interacting dyes 
	photobleaches by around $14 \times 10^6$ ADU/s. We note in (b) that the 
	brightness again reduces by a similar amount even though six 
	fluorophores in the fluorocube photobleach at the same time. This 
	dampened reduction in brightness may lead to significant undercounting 
	of subunits of a biomolecular complex under investigation.}
\label{photobleaching_traces}
\end{figure*}

\textit{Effects of Non-additivity.} Photobleaching occurs when fluorophores are 
chemically deactivated upon repeated excitations. This phenomenon finds applications in event counting experiments where the brightness decreases in a step-like pattern \cite{doi:10.1091/mbc.e14-06-1146} as fluorophores successively photobleach.  

In typical experiments, the accurate determination of the number and size of the steps is affected \cite{LIESCHE20152352} by instrumental noise, high numbers of fluorophores, overlapping photobleaching events, spatially varying intensity profile, and interactions between fluorophores and the surrounding molecules in the sample.  

In the absence of interactions, the brightness should decrease by approximately the same amount every time a fluorophore photobleaches. However, interactions can affect many aspects of physics dictating photobleaching event counting, including brightness of fluorophores \cite{Niekamp716787, schroder_scheible_steiner_vogelsang_tinnefeld_2019}, the stability of the chemical bonds against repeated excitations \cite{Niekamp716787, schroder_scheible_steiner_vogelsang_tinnefeld_2019}, and lifetimes of the excited states (as demonstrated earlier). For instance, it has been observed that, in the case of ATTO647N fluorocubes, the brightness 
is significantly reduced by interactions \cite{Niekamp716787}, most likely due to quenching by homo-FRET \cite{GAUTIER20013000} type of mechanism typically reducing fluorescence lifetime as well. On the other hand, photostability of these fluorocubes improves dramatically and many fold more photons are emitted \cite{Niekamp716787} resulting in delayed photobleaching, one contributing factor for which may simply be the highly dampened emission rate per fluorophore, however, in some cases $\sim$ 43 fold more photons are emitted \cite{Niekamp716787} compared to single dyes which cannot be explained without a full quantum mechanical treatment.

The effects of interactions noted above may manifest themselves in the form of irregular steps sizes (non-additive brightness) in photobleaching traces. This raises doubt on the assumption currently underlying all step counting methods relying on equal {\it average} step sizes (albeit varying around the average on account on photon shot noise and stochastic detector output) \cite{doi:10.1091/mbc.e14-06-1146}.

To illustrate non-additivity effects in step counting, we generate synthetic photobleaching traces using the recipe in \citet{Bryan2020.09.28.317057}. We generate photobleaching traces for a collection of interacting and non-interacting fluorophores. The region of interest (ROI) under study contains one interacting cluster of six fluorophores (a fluorocube) and another cluster of two sparsely distributed non-interacting fluorophores.

The algorithm used here for generating photobleaching traces requires us to 
specify brightness values for the fluorescent probes (isolated dyes and 
fluorocubes), which can be computed from the excitation probabilities as shown 
earlier. As the brightness values recorded by a camera are typically noisy, we 
model such measurements made by an EMCCD camera with a Gamma distribution 
\cite{Bryan2020.09.28.317057},
\be 
\mu_d \sim \mathbf{Gamma}(\mu_b/2+\mu_e, \,\, 2 G),
\label{camera_model}
\ee
where $\mu_e$ is the number of photons emitted in a second by the emitters, $\mu_b$ is the number of background photons emitted in a second, $\mu_d$ is the number of ADUs recorded by the detector in the same time interval, and $G$ is the camera gain, respectively.

From probability of excitation, we can compute the mean $\mu_e$ used in Eq.~\ref{camera_model} for both independent fluorophores and fluorocubes as $\pi_i/\Delta$. We recover that fluorocubes and independent fluorophores have roughly the same brightness (consistent with supplementary Figs.~3, 5, and 8 of \citet{Niekamp716787}).

More concretely, to compute the brightness in our experiments with 80 \% 
maximum laser power and three probes of the same type in the field of view, 
both single-dye cubes and fluorocubes have similar effective excitation 
probabilities of around 0.14 and 0.10 per pulse per probe (see 
Fig.~\ref{distributions}), respectively. With an inter-pulse time, $\Delta$, of 
around 100 ns, this corresponds to an emission rate, $\mu_{e}$, of around $1.4 
\times 10^6$ and $1.0 \times 10^6$ photons per second per probe. Additionally, 
the probabilities for background emissions, ${\pi}_{b}$, in both cases are 
similar and equal to approximately 0.10 and 0.12, which are equivalent to 
background emission rates, $\mu_{b}$, of $1.0 \times 10^6$ and $1.2 \times 
10^6$ photons/s. To generate the traces using the model from 
Eq.~\ref{camera_model}, we need to specify a camera gain and set it to 10 (a 
value, for instance, similar to that used by \citet{Bryan2020.09.28.317057}).

Fig.~\ref{photobleaching_traces}, illustrates what can be expected to happen when a single fluorophore (not interacting with any other) photobleaches (panel a) as compared to when an entire fluorocube photobleaches (panel b).

As is apparent from Fig.~\ref{photobleaching_traces}, the dampened brightness of the fluorocube has immediate implications: the step size cannot be considered an independent and identically distributed random variable for each fluorophore and may strongly depend on its local environment.

\textit{Conclusion.} Interactions among fluorophores and the surrounding chemical environment play a consequential role in the quantitative analysis of fluorescence imaging data. From our preliminary analysis of the photon arrival data for DNA origami structures labeled with ATTO647N dyes, we observe significant changes in the excitation probabilities and escape rates (lifetimes) of the excited states when interactions are involved. Interactions are also observed to significantly dampen the brightness of an interacting cluster of identical flurophores. The DNA scaffolding itself contributes to modification of the dye photophysics \cite{Niekamp716787}. These changes play an important role in the analysis of photobleaching event counting methods \cite{doi:10.1091/mbc.e14-06-1146}, where brightness traces for samples with large number of densely packed fluorophores cannot be modeled accurately within the existing non-fluorophore-interaction paradigm.

What is more, this finding also  suggests that determination of a protein complex's stoichiometry using PALM \cite{doi:10.1126/science.1127344, Sengupta2012, 
Lee17436} or photobleaching event counting \cite{doi:10.1091/mbc.e14-06-1146} or related methods \cite{Xie2018} may be impacted by the local distance between complex subunits.

So far, we have demonstrated that distances between fluorophores have an effect 
on the heights and lengths of steps in a photobleaching trace. However, in 
principle, assuming that we know the number of molecules, we can begin 
quantifying inter-molecular distances from accurate lifetime measurements in a 
manner similar to FRET \cite{10.3389/fphy.2019.00100} as well as from changes 
in emission frequencies. 

More precisely, we may design a means by which to precisely learn the 
interaction Hamiltonian, $H_{int}$,  in Eq.~\ref{hamiltonian} (and parameters 
such as escape rates, excitation probabilities, and photobleaching times) as a 
function of fluorophore separation, orientiation, and dipolar coupling. This 
calls for further experiments into precisely characterizing the relationship 
between photophysics and spatial distributions of identical fluorophores.

\textit{Experimental Methods.} 
We acquired data on an SP8 point-scanning system (Leica) with a Picoquant TCSPC module and excitation delivered by a WLL at a 10 MHz repetition rate. The emission spectrum was split onto two SMD HyD detectors: 581-603 and 608-687 nm for 561 nm excitation, or 646-670 and 675-752 nm for 640 nm excitation. We used an HC PL APO CS2 100x/1.4NA objective, a pinhole size of 1AU, 9.5 nm pixels, and a dwell-time of 97.66 microseconds. Data was extracted from .ptu files using a parser written in python. 

We assembled the large cubes \cite{compact_DNA} and the fluorocubes as 
described by \citet{Niekamp716787}. Briefly, for each six-dye fluorocube and 
single-dye cube we used four 32 bp long oligonucleotide strands, each modified 
either with dyes or biotin (sequences shown in supplementary Table 1). We mixed 
each of the four oligos (for the fluorocubes) or the 28 oligos (for the large 
cubes, sequences shown in supplementary Table 1) to a final concentration of 10 
$\mu$M in folding buffer (5 mM Tris pH 8.5, 1 mM EDTA and 40 mM MgCl$_2$). We 
then annealed the oligos by denaturation at 85 $^{\circ}$C for 5 min, followed 
by cooling from 80 to 65 $^{\circ}$ C with a decrease of 1 $^{\circ}$C per 5 
min, followed by further cooling from 65 to 25 $^{\circ}$ C with a decrease of 
1 $^{\circ}$C per 20 min. Afterwards the samples were held at 4 $^{\circ}$C. We 
then analyzed the folding products by 3.0$\%$ agarose gel electrophoresis in 
TBE (45 mM Tris-borate and 1 mM EDTA) with 12 mM MgCl$_2$ at 70 V for 2.5 h on 
ice. We finally purified the samples by extraction and centrifugation in 
Freeze’N Squeeze columns (Bio-Rad Sciences, 732-6165). 

We prepared flow-cells as described by \citet{Niekamp716787}. First, we used a laser cutter to cut custom three-cell flow chambers out of double-sided adhesive sheets (Soles2dance, 9474-08x12 - 3M 9474LE 300LSE). We cleaned 170 $\mu$m thick coverslips (Zeiss, 474030-9000-000) in a 5$\%$ v/v solution of Hellmanex III (Sigma, Z805939-1EA) at 50 $^{\circ}$ C overnight and washed extensively with Milli-Q water afterwards. Then, we used three-cell flow chambers together with glass slides (Thermo Fisher Scientific, 12-550-123) and coverslips to assemble the flow-cells.

The preparation method for flow-cells is the same for the fluorocubes and the 
large cubes \cite{Niekamp716787}. Briefly, we first added 10 $\mu$l of 5 mg/ml 
Biotin-BSA (Thermo Scientific, 29130) in BRB80 (80 mM Pipes (pH 6.8), 1 mM 
MgCl$_2$, 1 mM EGTA) to the flow-cell and incubated for 2 min. We then added an 
additional 10 $\mu$l of 5 mg/ml Biotin-BSA in BRB80 and incubated for another 2 
min. Afterwards, we the flow-cell was washed with 20 $\mu$l of fluorocube 
buffer (20 mM Tris pH 8.0, 1 mM EDTA, 20 mM Mg-Ac and 50 mM NaCl) with 2 mg/ml 
of $\beta$-casein (Sigma, C6905) and 0.4 mg/ml $\kappa$-casein (Sigma, C0406).  
Next, we added 10 $\mu$l of 0.5 mg/ml Streptavidin (Vector Laboratories, 
SA-5000) in PBS (pH 7.4), incubated for 2 min and then washed with 20 $\mu$l of 
fluorocube buffer with 2 mg/ml $\beta$-casein and 0.4 mg/ml $\kappa$-casein.  
Afterwards, we added either fluorocubes or large cubes in fluorocube buffer 
with 2 mg/ml $\beta$-casein and 0.4 mg/ml $\kappa$-casein and incubated for 5 
min. Next, the flow-cell was washed with 30 $\mu$l of fluorocube buffer with 2 
mg/ml $\beta$-casein and 0.4 mg/ml $\kappa$-casein. Finally, we added the 
protocatechuic acid (PCA) / protocatechuate-3,4-dioxygenase (PCD) / Trolox 
oxygen scavenging system (3, 4) in fluorocube buffer with 2 mg/ml 
$\beta$-casein, and 0.4 mg/ml $\kappa$-casein to the flow-cell. For the 
PCA/PCD/Trolox oxygen scavenging system we used 2.5 mM of PCA (Sigma, 37580) at 
pH 9.0, 5 U of PCD (Oriental Yeast Company Americas Inc., 46852004), and 1 mM 
Trolox (Sigma, 238813) at pH 9.5.

\begin{acknowledgement}
We thank Dr Andrew York, Dr Nico Stuurman, and Dr Maria Ingaramo for providing the experimental data, and for regular feedback and discussions. We also thank Dr Douglas Shepherd for setting up the collaboration and providing insight into the workings of detectors and other experimental equipment. S. P. acknowledges support from the NIH NIGMS (R01GM130745) for supporting early efforts in nonparametrics and NIH NIGMS (R01GM134426) for supporting single-photon efforts. 
\end{acknowledgement}




\bibliography{manuscript}

\providecommand{\latin}[1]{#1}
\makeatletter
\providecommand{\doi}
  {\begingroup\let\do\@makeother\dospecials
  \catcode`\{=1 \catcode`\}=2 \doi@aux}
\providecommand{\doi@aux}[1]{\endgroup\texttt{#1}}
\makeatother
\providecommand*\mcitethebibliography{\thebibliography}
\csname @ifundefined\endcsname{endmcitethebibliography}
  {\let\endmcitethebibliography\endthebibliography}{}
\begin{mcitethebibliography}{59}
\providecommand*\natexlab[1]{#1}
\providecommand*\mciteSetBstSublistMode[1]{}
\providecommand*\mciteSetBstMaxWidthForm[2]{}
\providecommand*\mciteBstWouldAddEndPuncttrue
  {\def\EndOfBibitem{\unskip.}}
\providecommand*\mciteBstWouldAddEndPunctfalse
  {\let\EndOfBibitem\relax}
\providecommand*\mciteSetBstMidEndSepPunct[3]{}
\providecommand*\mciteSetBstSublistLabelBeginEnd[3]{}
\providecommand*\EndOfBibitem{}
\mciteSetBstSublistMode{f}
\mciteSetBstMaxWidthForm{subitem}{(\alph{mcitesubitemcount})}
\mciteSetBstSublistLabelBeginEnd
  {\mcitemaxwidthsubitemform\space}
  {\relax}
  {\relax}

\bibitem[Shashkova and Leake(2017)Shashkova, and Leake]{10.1042/BSR20170031}
Shashkova,~S.; Leake,~M. {Single-molecule fluorescence microscopy review:
  shedding new light on old problems}. \emph{Bioscience Reports} \textbf{2017},
  \emph{37}, BSR20170031\relax
\mciteBstWouldAddEndPuncttrue
\mciteSetBstMidEndSepPunct{\mcitedefaultmidpunct}
{\mcitedefaultendpunct}{\mcitedefaultseppunct}\relax
\EndOfBibitem
\bibitem[Betzig \latin{et~al.}(2006)Betzig, Patterson, Sougrat, Lindwasser,
  Olenych, Bonifacino, Davidson, Lippincott-Schwartz, and
  Hess]{doi:10.1126/science.1127344}
Betzig,~E.; Patterson,~G.~H.; Sougrat,~R.; Lindwasser,~O.~W.; Olenych,~S.;
  Bonifacino,~J.~S.; Davidson,~M.~W.; Lippincott-Schwartz,~J.; Hess,~H.~F.
  Imaging Intracellular Fluorescent Proteins at Nanometer Resolution.
  \emph{Science} \textbf{2006}, \emph{313}, 1642--1645\relax
\mciteBstWouldAddEndPuncttrue
\mciteSetBstMidEndSepPunct{\mcitedefaultmidpunct}
{\mcitedefaultendpunct}{\mcitedefaultseppunct}\relax
\EndOfBibitem
\bibitem[Hell and Wichmann(1994)Hell, and Wichmann]{Hell:94}
Hell,~S.~W.; Wichmann,~J. Breaking the diffraction resolution limit by
  stimulated emission: stimulated-emission-depletion fluorescence microscopy.
  \emph{Opt. Lett.} \textbf{1994}, \emph{19}, 780--782\relax
\mciteBstWouldAddEndPuncttrue
\mciteSetBstMidEndSepPunct{\mcitedefaultmidpunct}
{\mcitedefaultendpunct}{\mcitedefaultseppunct}\relax
\EndOfBibitem
\bibitem[Tsekouras \latin{et~al.}(2016)Tsekouras, Custer, Jashnsaz, Walter, and
  Pressé]{Tsekouras2016}
Tsekouras,~K.; Custer,~T.~C.; Jashnsaz,~H.; Walter,~N.~G.; Pressé,~S. A novel
  method to accurately locate and count large numbers of steps by
  photobleaching. \emph{Molecular Biology of the Cell} \textbf{2016},
  \emph{27}, 3601--3615, PMID: 27654946\relax
\mciteBstWouldAddEndPuncttrue
\mciteSetBstMidEndSepPunct{\mcitedefaultmidpunct}
{\mcitedefaultendpunct}{\mcitedefaultseppunct}\relax
\EndOfBibitem
\bibitem[Bryan \latin{et~al.}(2020)Bryan, Sgouralis, and
  Press{\'e}]{Bryan2020.09.28.317057}
Bryan,~J.~S.; Sgouralis,~I.; Press{\'e},~S. Enumerating High Numbers of
  Fluorophores from Photobleaching Experiments: a Bayesian Nonparametrics
  Approach. \emph{bioRxiv} \textbf{2020}, \relax
\mciteBstWouldAddEndPunctfalse
\mciteSetBstMidEndSepPunct{\mcitedefaultmidpunct}
{}{\mcitedefaultseppunct}\relax
\EndOfBibitem
\bibitem[Arant and Ulbrich(2014)Arant, and Ulbrich]{Arant2014}
Arant,~R.~J.; Ulbrich,~M.~H. Deciphering the Subunit Composition of Multimeric
  Proteins by Counting Photobleaching Steps. \emph{ChemPhysChem} \textbf{2014},
  \emph{15}, 600--605\relax
\mciteBstWouldAddEndPuncttrue
\mciteSetBstMidEndSepPunct{\mcitedefaultmidpunct}
{\mcitedefaultendpunct}{\mcitedefaultseppunct}\relax
\EndOfBibitem
\bibitem[Ulbrich and Isacoff(2007)Ulbrich, and Isacoff]{Ulbrich2007}
Ulbrich,~M.~H.; Isacoff,~E.~Y. Subunit counting in membrane-bound proteins.
  \emph{Nature Methods} \textbf{2007}, \emph{4}, 319--321\relax
\mciteBstWouldAddEndPuncttrue
\mciteSetBstMidEndSepPunct{\mcitedefaultmidpunct}
{\mcitedefaultendpunct}{\mcitedefaultseppunct}\relax
\EndOfBibitem
\bibitem[Chen \latin{et~al.}(2014)Chen, Deffenbaugh, Anderson, and
  Hancock]{doi:10.1091/mbc.e14-06-1146}
Chen,~Y.; Deffenbaugh,~N.~C.; Anderson,~C.~T.; Hancock,~W.~O. Molecular
  counting by photobleaching in protein complexes with many subunits: best
  practices and application to the cellulose synthesis complex. \emph{Molecular
  Biology of the Cell} \textbf{2014}, \emph{25}, 3630--3642, PMID:
  25232006\relax
\mciteBstWouldAddEndPuncttrue
\mciteSetBstMidEndSepPunct{\mcitedefaultmidpunct}
{\mcitedefaultendpunct}{\mcitedefaultseppunct}\relax
\EndOfBibitem
\bibitem[Delalez \latin{et~al.}(2010)Delalez, Wadhams, Rosser, Xue, Brown,
  Dobbie, Berry, Leake, and Armitage]{Delalez11347}
Delalez,~N.~J.; Wadhams,~G.~H.; Rosser,~G.; Xue,~Q.; Brown,~M.~T.;
  Dobbie,~I.~M.; Berry,~R.~M.; Leake,~M.~C.; Armitage,~J.~P. Signal-dependent
  turnover of the bacterial flagellar switch protein FliM. \emph{Proceedings of
  the National Academy of Sciences} \textbf{2010}, \emph{107},
  11347--11351\relax
\mciteBstWouldAddEndPuncttrue
\mciteSetBstMidEndSepPunct{\mcitedefaultmidpunct}
{\mcitedefaultendpunct}{\mcitedefaultseppunct}\relax
\EndOfBibitem
\bibitem[Engel \latin{et~al.}(2009)Engel, Ludington, and
  Marshall]{10.1083/jcb.200812084}
Engel,~B.~D.; Ludington,~W.~B.; Marshall,~W.~F. {Intraflagellar transport
  particle size scales inversely with flagellar length: revisiting the
  balance-point length control model}. \emph{Journal of Cell Biology}
  \textbf{2009}, \emph{187}, 81--89\relax
\mciteBstWouldAddEndPuncttrue
\mciteSetBstMidEndSepPunct{\mcitedefaultmidpunct}
{\mcitedefaultendpunct}{\mcitedefaultseppunct}\relax
\EndOfBibitem
\bibitem[Lawrimore \latin{et~al.}(2011)Lawrimore, Bloom, and
  Salmon]{10.1083/jcb.201106036}
Lawrimore,~J.; Bloom,~K.~S.; Salmon,~E. {Point centromeres contain more than a
  single centromere-specific Cse4 (CENP-A) nucleosome}. \emph{Journal of Cell
  Biology} \textbf{2011}, \emph{195}, 573--582\relax
\mciteBstWouldAddEndPuncttrue
\mciteSetBstMidEndSepPunct{\mcitedefaultmidpunct}
{\mcitedefaultendpunct}{\mcitedefaultseppunct}\relax
\EndOfBibitem
\bibitem[Coffman \latin{et~al.}(2011)Coffman, Wu, Parthun, and
  Wu]{10.1083/jcb.201106078}
Coffman,~V.~C.; Wu,~P.; Parthun,~M.~R.; Wu,~J.-Q. {CENP-A exceeds microtubule
  attachment sites in centromere clusters of both budding and fission yeast}.
  \emph{Journal of Cell Biology} \textbf{2011}, \emph{195}, 563--572\relax
\mciteBstWouldAddEndPuncttrue
\mciteSetBstMidEndSepPunct{\mcitedefaultmidpunct}
{\mcitedefaultendpunct}{\mcitedefaultseppunct}\relax
\EndOfBibitem
\bibitem[McGuire \latin{et~al.}(2012)McGuire, Aurousseau, Bowie, and
  Blunck]{McGuire2012}
McGuire,~H.; Aurousseau,~M. R.~P.; Bowie,~D.; Blunck,~R. Automating single
  subunit counting of membrane proteins in mammalian cells. \emph{The Journal
  of Biological Chemistry} \textbf{2012}, \emph{287}, 35912--35921\relax
\mciteBstWouldAddEndPuncttrue
\mciteSetBstMidEndSepPunct{\mcitedefaultmidpunct}
{\mcitedefaultendpunct}{\mcitedefaultseppunct}\relax
\EndOfBibitem
\bibitem[Demuro \latin{et~al.}(2011)Demuro, Penna, Safrina, Yeromin,
  Amcheslavsky, Cahalan, and Parker]{Demuro17832}
Demuro,~A.; Penna,~A.; Safrina,~O.; Yeromin,~A.~V.; Amcheslavsky,~A.;
  Cahalan,~M.~D.; Parker,~I. Subunit stoichiometry of human Orai1 and Orai3
  channels in closed and open states. \emph{Proceedings of the National Academy
  of Sciences} \textbf{2011}, \emph{108}, 17832--17837\relax
\mciteBstWouldAddEndPuncttrue
\mciteSetBstMidEndSepPunct{\mcitedefaultmidpunct}
{\mcitedefaultendpunct}{\mcitedefaultseppunct}\relax
\EndOfBibitem
\bibitem[Hastie \latin{et~al.}(2013)Hastie, Ulbrich, Wang, Arant, Lau, Zhang,
  Isacoff, and Chen]{Hastie5163}
Hastie,~P.; Ulbrich,~M.~H.; Wang,~H.-L.; Arant,~R.~J.; Lau,~A.~G.; Zhang,~Z.;
  Isacoff,~E.~Y.; Chen,~L. AMPA receptor/TARP stoichiometry visualized by
  single-molecule subunit counting. \emph{Proceedings of the National Academy
  of Sciences} \textbf{2013}, \emph{110}, 5163--5168\relax
\mciteBstWouldAddEndPuncttrue
\mciteSetBstMidEndSepPunct{\mcitedefaultmidpunct}
{\mcitedefaultendpunct}{\mcitedefaultseppunct}\relax
\EndOfBibitem
\bibitem[Arumugam \latin{et~al.}(2012)Arumugam, Lee, and
  Benkovic]{Arumugam2009}
Arumugam,~S.~R.; Lee,~T.-H.; Benkovic,~S.~J. Investigation of Stoichiometry of
  T4 Bacteriophage Helicase Loader Protein (gp59). \emph{The Journal of
  Biological Chemistry} \textbf{2012}, \emph{284}, 29283--29289\relax
\mciteBstWouldAddEndPuncttrue
\mciteSetBstMidEndSepPunct{\mcitedefaultmidpunct}
{\mcitedefaultendpunct}{\mcitedefaultseppunct}\relax
\EndOfBibitem
\bibitem[Llorente-Garcia \latin{et~al.}(2014)Llorente-Garcia, Lenn, Erhardt,
  Harriman, Liu, Robson, Chiu, Matthews, Willis, Bray, Lee, Shin, Bustamante,
  Liphardt, Friedrich, Mullineaux, and Leake]{LLORENTEGARCIA2014811}
Llorente-Garcia,~I.; Lenn,~T.; Erhardt,~H.; Harriman,~O.~L.; Liu,~L.-N.;
  Robson,~A.; Chiu,~S.-W.; Matthews,~S.; Willis,~N.~J.; Bray,~C.~D.
  \latin{et~al.}  Single-molecule in vivo imaging of bacterial respiratory
  complexes indicates delocalized oxidative phosphorylation. \emph{Biochimica
  et Biophysica Acta (BBA) - Bioenergetics} \textbf{2014}, \emph{1837},
  811--824\relax
\mciteBstWouldAddEndPuncttrue
\mciteSetBstMidEndSepPunct{\mcitedefaultmidpunct}
{\mcitedefaultendpunct}{\mcitedefaultseppunct}\relax
\EndOfBibitem
\bibitem[Pitchiaya \latin{et~al.}(2013)Pitchiaya, Krishnan, Custer, and
  Walter]{PITCHIAYA2013188}
Pitchiaya,~S.; Krishnan,~V.; Custer,~T.~C.; Walter,~N.~G. Dissecting non-coding
  RNA mechanisms in cellulo by Single-molecule High-Resolution Localization and
  Counting. \emph{Methods} \textbf{2013}, \emph{63}, 188--199, Non-coding RNA
  Methods\relax
\mciteBstWouldAddEndPuncttrue
\mciteSetBstMidEndSepPunct{\mcitedefaultmidpunct}
{\mcitedefaultendpunct}{\mcitedefaultseppunct}\relax
\EndOfBibitem
\bibitem[Pitchiaya \latin{et~al.}(2014)Pitchiaya, Heinicke, Custer, and
  Walter]{pitchiaya2014}
Pitchiaya,~S.; Heinicke,~L.~A.; Custer,~T.~C.; Walter,~N.~G. Single Molecule
  Fluorescence Approaches Shed Light on Intracellular RNAs. \emph{Chemical
  Reviews} \textbf{2014}, \emph{114}, 3224–3265\relax
\mciteBstWouldAddEndPuncttrue
\mciteSetBstMidEndSepPunct{\mcitedefaultmidpunct}
{\mcitedefaultendpunct}{\mcitedefaultseppunct}\relax
\EndOfBibitem
\bibitem[Shu \latin{et~al.}(2007)Shu, Zhang, Jin, and Guo]{shu2007}
Shu,~D.; Zhang,~H.; Jin,~J.; Guo,~P. Counting of six pRNAs of phi29
  DNA-packaging motor with customized single-molecule dual-view system.
  \emph{The EMBO Journal} \textbf{2007}, \emph{26}, 527--537\relax
\mciteBstWouldAddEndPuncttrue
\mciteSetBstMidEndSepPunct{\mcitedefaultmidpunct}
{\mcitedefaultendpunct}{\mcitedefaultseppunct}\relax
\EndOfBibitem
\bibitem[Leake \latin{et~al.}(2006)Leake, Chandler, Wadhams, Bai, Berry, and
  Armitage]{Leake2006}
Leake,~M.~C.; Chandler,~J.~H.; Wadhams,~G.~H.; Bai,~F.; Berry,~R.~M.;
  Armitage,~J.~P. Stoichiometry and turnover in single, functioning membrane
  protein complexes. \emph{Nature} \textbf{2006}, \emph{443}, 355--358\relax
\mciteBstWouldAddEndPuncttrue
\mciteSetBstMidEndSepPunct{\mcitedefaultmidpunct}
{\mcitedefaultendpunct}{\mcitedefaultseppunct}\relax
\EndOfBibitem
\bibitem[Das \latin{et~al.}(2007)Das, Darshi, Cheley, Wallace, and
  Bayley]{Das2007}
Das,~S.~K.; Darshi,~M.; Cheley,~S.; Wallace,~M.~I.; Bayley,~H. Membrane Protein
  Stoichiometry Determined from the Step-Wise Photobleaching of Dye-Labelled
  Subunits. \emph{ChemBioChem} \textbf{2007}, \emph{8}, 994--999\relax
\mciteBstWouldAddEndPuncttrue
\mciteSetBstMidEndSepPunct{\mcitedefaultmidpunct}
{\mcitedefaultendpunct}{\mcitedefaultseppunct}\relax
\EndOfBibitem
\bibitem[Singh \latin{et~al.}(2020)Singh, Van~Slyke, Sirenko, Song,
  Kammermeier, and Zipfel]{Singh2020}
Singh,~A.; Van~Slyke,~A.~L.; Sirenko,~M.; Song,~A.; Kammermeier,~P.~J.;
  Zipfel,~W.~R. Stoichiometric analysis of protein complexes by cell fusion and
  single molecule imaging. \emph{Scientific Reports} \textbf{2020}, \emph{10},
  14866\relax
\mciteBstWouldAddEndPuncttrue
\mciteSetBstMidEndSepPunct{\mcitedefaultmidpunct}
{\mcitedefaultendpunct}{\mcitedefaultseppunct}\relax
\EndOfBibitem
\bibitem[Dhara \latin{et~al.}(2020)Dhara, Sadhukhan, Sheetz, Olsson,
  Raghavachari, and Flood]{Dhara2020}
Dhara,~A.; Sadhukhan,~T.; Sheetz,~E.~G.; Olsson,~A.~H.; Raghavachari,~K.;
  Flood,~A.~H. Zero-Overlap Fluorophores for Fluorescent Studies at Any
  Concentration. \emph{Journal of the American Chemical Society} \textbf{2020},
  \emph{142}, 12167--12180\relax
\mciteBstWouldAddEndPuncttrue
\mciteSetBstMidEndSepPunct{\mcitedefaultmidpunct}
{\mcitedefaultendpunct}{\mcitedefaultseppunct}\relax
\EndOfBibitem
\bibitem[Sengupta and Lippincott-Schwartz(2012)Sengupta, and
  Lippincott-Schwartz]{Sengupta2012}
Sengupta,~P.; Lippincott-Schwartz,~J. Quantitative analysis of photoactivated
  localization microscopy (PALM) datasets using pair-correlation analysis.
  \emph{BioEssays} \textbf{2012}, \emph{34}, 396--405\relax
\mciteBstWouldAddEndPuncttrue
\mciteSetBstMidEndSepPunct{\mcitedefaultmidpunct}
{\mcitedefaultendpunct}{\mcitedefaultseppunct}\relax
\EndOfBibitem
\bibitem[Lee \latin{et~al.}(2012)Lee, Shin, Lee, and Bustamante]{Lee17436}
Lee,~S.-H.; Shin,~J.~Y.; Lee,~A.; Bustamante,~C. Counting single
  photoactivatable fluorescent molecules by photoactivated localization
  microscopy (PALM). \emph{Proceedings of the National Academy of Sciences}
  \textbf{2012}, \emph{109}, 17436--17441\relax
\mciteBstWouldAddEndPuncttrue
\mciteSetBstMidEndSepPunct{\mcitedefaultmidpunct}
{\mcitedefaultendpunct}{\mcitedefaultseppunct}\relax
\EndOfBibitem
\bibitem[Rush \latin{et~al.}(2006)Rush, Bates, and Zhuang]{STORM}
Rush,~M.~J.; Bates,~M.; Zhuang,~X. Sub-diffraction-limit imaging by stochastic
  optical reconstruction microscopy (STORM). \emph{Nature Methods}
  \textbf{2006}, \emph{3}, 793--796\relax
\mciteBstWouldAddEndPuncttrue
\mciteSetBstMidEndSepPunct{\mcitedefaultmidpunct}
{\mcitedefaultendpunct}{\mcitedefaultseppunct}\relax
\EndOfBibitem
\bibitem[Amann and Fuchs(2008)Amann, and Fuchs]{Amann2008}
Amann,~R.; Fuchs,~B.~M. Single-cell identification in microbial communities by
  improved fluorescence in situ hybridization techniques. \emph{Nature Reviews
  Microbiology} \textbf{2008}, \emph{6}, 339--348\relax
\mciteBstWouldAddEndPuncttrue
\mciteSetBstMidEndSepPunct{\mcitedefaultmidpunct}
{\mcitedefaultendpunct}{\mcitedefaultseppunct}\relax
\EndOfBibitem
\bibitem[Grubmayer \latin{et~al.}(2019)Grubmayer, Yserentant, and
  Herten]{Gru_mayer_2019}
Grubmayer,~K.~S.; Yserentant,~K.; Herten,~D.-P. Photons in - numbers out:
  perspectives in quantitative fluorescence microscopy for in situ protein
  counting. \emph{Methods and Applications in Fluorescence} \textbf{2019},
  \emph{7}, 012003\relax
\mciteBstWouldAddEndPuncttrue
\mciteSetBstMidEndSepPunct{\mcitedefaultmidpunct}
{\mcitedefaultendpunct}{\mcitedefaultseppunct}\relax
\EndOfBibitem
\bibitem[Dankovich and Rizzoli(2021)Dankovich, and
  Rizzoli]{DANKOVICH2021102134}
Dankovich,~T.~M.; Rizzoli,~S.~O. Challenges facing quantitative large-scale
  optical super-resolution, and some simple solutions. \emph{iScience}
  \textbf{2021}, \emph{24}, 102134\relax
\mciteBstWouldAddEndPuncttrue
\mciteSetBstMidEndSepPunct{\mcitedefaultmidpunct}
{\mcitedefaultendpunct}{\mcitedefaultseppunct}\relax
\EndOfBibitem
\bibitem[Mandracchia \latin{et~al.}(2020)Mandracchia, Hua, Guo, Son, and
  Jia]{Mandracchia2020}
Mandracchia,~B.; Hua,~X.; Guo,~C.; Son,~J.; Jia,~T. U. .~S. Fast and accurate
  sCMOS noise correction for fluorescence microscopy. \emph{Nature
  Communications} \textbf{2020}, \emph{11}, 94--\relax
\mciteBstWouldAddEndPuncttrue
\mciteSetBstMidEndSepPunct{\mcitedefaultmidpunct}
{\mcitedefaultendpunct}{\mcitedefaultseppunct}\relax
\EndOfBibitem
\bibitem[Hirsch \latin{et~al.}(2013)Hirsch, Wareham, Martin-Fernandez, Hobson,
  and Rolfe]{emCCD2013}
Hirsch,~M.; Wareham,~R.~J.; Martin-Fernandez,~M.~L.; Hobson,~M.~P.;
  Rolfe,~D.~J. A Stochastic Model for Electron Multiplication Charge-Coupled
  Devices – From Theory to Practice. \emph{PLOS ONE} \textbf{2013}, \emph{8},
  1--13\relax
\mciteBstWouldAddEndPuncttrue
\mciteSetBstMidEndSepPunct{\mcitedefaultmidpunct}
{\mcitedefaultendpunct}{\mcitedefaultseppunct}\relax
\EndOfBibitem
\bibitem[Luchowski \latin{et~al.}(2009)Luchowski, Gryczynski, Sarkar, Borejdo,
  Szabelski, Kapusta, and Gryczynski]{doi:10.1063/1.3095677}
Luchowski,~R.; Gryczynski,~Z.; Sarkar,~P.; Borejdo,~J.; Szabelski,~M.;
  Kapusta,~P.; Gryczynski,~I. Instrument response standard in time-resolved
  fluorescence. \emph{Review of Scientific Instruments} \textbf{2009},
  \emph{80}, 033109\relax
\mciteBstWouldAddEndPuncttrue
\mciteSetBstMidEndSepPunct{\mcitedefaultmidpunct}
{\mcitedefaultendpunct}{\mcitedefaultseppunct}\relax
\EndOfBibitem
\bibitem[Szabelski \latin{et~al.}(2009)Szabelski, Luchowski, Gryczynski,
  Kapusta, Ortmann, and Gryczynski]{SZABELSKI2009153}
Szabelski,~M.; Luchowski,~R.; Gryczynski,~Z.; Kapusta,~P.; Ortmann,~U.;
  Gryczynski,~I. Evaluation of instrument response functions for lifetime
  imaging detectors using quenched Rose Bengal solutions. \emph{Chemical
  Physics Letters} \textbf{2009}, \emph{471}, 153--159\relax
\mciteBstWouldAddEndPuncttrue
\mciteSetBstMidEndSepPunct{\mcitedefaultmidpunct}
{\mcitedefaultendpunct}{\mcitedefaultseppunct}\relax
\EndOfBibitem
\bibitem[Lakowicz(2013)]{lakowicz2013principles}
Lakowicz,~J.~R. \emph{Principles of fluorescence spectroscopy}; Springer
  science \& business media, 2013\relax
\mciteBstWouldAddEndPuncttrue
\mciteSetBstMidEndSepPunct{\mcitedefaultmidpunct}
{\mcitedefaultendpunct}{\mcitedefaultseppunct}\relax
\EndOfBibitem
\bibitem[Sgouralis \latin{et~al.}(2019)Sgouralis, Madaan, Djutanta, Kha,
  Hariadi, and Pressé]{doi:10.1021/acs.jpcb.8b09752}
Sgouralis,~I.; Madaan,~S.; Djutanta,~F.; Kha,~R.; Hariadi,~R.~F.; Pressé,~S. A
  Bayesian Nonparametric Approach to Single Molecule Förster Resonance Energy
  Transfer. \emph{The Journal of Physical Chemistry B} \textbf{2019},
  \emph{123}, 675--688\relax
\mciteBstWouldAddEndPuncttrue
\mciteSetBstMidEndSepPunct{\mcitedefaultmidpunct}
{\mcitedefaultendpunct}{\mcitedefaultseppunct}\relax
\EndOfBibitem
\bibitem[Coffman and Wu(2012)Coffman, and Wu]{Coffman2012}
Coffman,~V.~C.; Wu,~J.-Q. Counting protein molecules using quantitative
  fluorescence microscopy. \emph{Trends in Biochemical Sciences} \textbf{2012},
  \emph{37}, 499--506\relax
\mciteBstWouldAddEndPuncttrue
\mciteSetBstMidEndSepPunct{\mcitedefaultmidpunct}
{\mcitedefaultendpunct}{\mcitedefaultseppunct}\relax
\EndOfBibitem
\bibitem[Xie \latin{et~al.}(2018)Xie, Timme, and Wood]{Xie2018}
Xie,~F.; Timme,~K.~A.; Wood,~J.~R. Using Single Molecule mRNA Fluorescent in
  Situ Hybridization (RNA-FISH) to Quantify mRNAs in Individual Murine Oocytes
  and Embryos. \emph{Scientific Reports} \textbf{2018}, \emph{8}, 7930\relax
\mciteBstWouldAddEndPuncttrue
\mciteSetBstMidEndSepPunct{\mcitedefaultmidpunct}
{\mcitedefaultendpunct}{\mcitedefaultseppunct}\relax
\EndOfBibitem
\bibitem[Niekamp \latin{et~al.}(2019)Niekamp, Stuurman, and
  Vale]{Niekamp716787}
Niekamp,~S.; Stuurman,~N.; Vale,~R.~D. A 6-nm ultra-photostable DNA Fluorocube
  for fluorescence imaging. \emph{bioRxiv} \textbf{2019}, \relax
\mciteBstWouldAddEndPunctfalse
\mciteSetBstMidEndSepPunct{\mcitedefaultmidpunct}
{}{\mcitedefaultseppunct}\relax
\EndOfBibitem
\bibitem[Schröder \latin{et~al.}(2019)Schröder, Scheible, Steiner, Vogelsang,
  and Tinnefeld]{schroder_scheible_steiner_vogelsang_tinnefeld_2019}
Schröder,~T.; Scheible,~M.~B.; Steiner,~F.; Vogelsang,~J.; Tinnefeld,~P.
  Interchromophoric Interactions Determine the Maximum Brightness Density in
  DNA Origami Structures. \emph{Nano Letters} \textbf{2019}, \emph{19},
  1275–1281\relax
\mciteBstWouldAddEndPuncttrue
\mciteSetBstMidEndSepPunct{\mcitedefaultmidpunct}
{\mcitedefaultendpunct}{\mcitedefaultseppunct}\relax
\EndOfBibitem
\bibitem[Helmerich \latin{et~al.}(2022)Helmerich, Beliu, Taban, Meub, Streit,
  Kuhlemann, Doose, and Sauer]{Helmerich2022.02.08.479592}
Helmerich,~D.~A.; Beliu,~G.; Taban,~D.; Meub,~M.; Streit,~M.; Kuhlemann,~A.;
  Doose,~S.; Sauer,~M. Sub-10 nm fluorescence imaging. \emph{bioRxiv}
  \textbf{2022}, \relax
\mciteBstWouldAddEndPunctfalse
\mciteSetBstMidEndSepPunct{\mcitedefaultmidpunct}
{}{\mcitedefaultseppunct}\relax
\EndOfBibitem
\bibitem[Lakowicz(2006)]{ref1}
Lakowicz,~J.~R., Ed. \emph{Principles of Fluorescence Spectroscopy}; Springer
  US: Boston, MA, 2006; pp 277--330\relax
\mciteBstWouldAddEndPuncttrue
\mciteSetBstMidEndSepPunct{\mcitedefaultmidpunct}
{\mcitedefaultendpunct}{\mcitedefaultseppunct}\relax
\EndOfBibitem
\bibitem[Jones and Bradshaw(2019)Jones, and Bradshaw]{10.3389/fphy.2019.00100}
Jones,~G.~A.; Bradshaw,~D.~S. Resonance Energy Transfer: From Fundamental
  Theory to Recent Applications. \emph{Frontiers in Physics} \textbf{2019},
  \emph{7}, 100\relax
\mciteBstWouldAddEndPuncttrue
\mciteSetBstMidEndSepPunct{\mcitedefaultmidpunct}
{\mcitedefaultendpunct}{\mcitedefaultseppunct}\relax
\EndOfBibitem
\bibitem[Griffiths and Schroeter(2018)Griffiths, and
  Schroeter]{griffiths_schroeter_2018}
Griffiths,~D.~J.; Schroeter,~D.~F. \emph{Introduction to Quantum Mechanics},
  3rd ed.; Cambridge University Press, 2018\relax
\mciteBstWouldAddEndPuncttrue
\mciteSetBstMidEndSepPunct{\mcitedefaultmidpunct}
{\mcitedefaultendpunct}{\mcitedefaultseppunct}\relax
\EndOfBibitem
\bibitem[Sgouralis and Pressé(2017)Sgouralis, and Pressé]{SGOURALIS20172117}
Sgouralis,~I.; Pressé,~S. ICON: An Adaptation of Infinite HMMs for Time Traces
  with Drift. \emph{Biophysical Journal} \textbf{2017}, \emph{112},
  2117--2126\relax
\mciteBstWouldAddEndPuncttrue
\mciteSetBstMidEndSepPunct{\mcitedefaultmidpunct}
{\mcitedefaultendpunct}{\mcitedefaultseppunct}\relax
\EndOfBibitem
\bibitem[Sgouralis and Pressé(2017)Sgouralis, and Pressé]{SGOURALIS20172021}
Sgouralis,~I.; Pressé,~S. An Introduction to Infinite HMMs for Single-Molecule
  Data Analysis. \emph{Biophysical Journal} \textbf{2017}, \emph{112},
  2021--2029\relax
\mciteBstWouldAddEndPuncttrue
\mciteSetBstMidEndSepPunct{\mcitedefaultmidpunct}
{\mcitedefaultendpunct}{\mcitedefaultseppunct}\relax
\EndOfBibitem
\bibitem[Jazani \latin{et~al.}(2019)Jazani, Sgouralis, Shafraz, Levitus,
  Sivasankar, and Press{\'e}]{Jazani2019}
Jazani,~S.; Sgouralis,~I.; Shafraz,~O.~M.; Levitus,~M.; Sivasankar,~S.;
  Press{\'e},~S. An alternative framework for fluorescence correlation
  spectroscopy. \emph{Nature Communications} \textbf{2019}, \emph{10},
  3662\relax
\mciteBstWouldAddEndPuncttrue
\mciteSetBstMidEndSepPunct{\mcitedefaultmidpunct}
{\mcitedefaultendpunct}{\mcitedefaultseppunct}\relax
\EndOfBibitem
\bibitem[Sgouralis \latin{et~al.}(2018)Sgouralis, Whitmore, Lapidus, Comstock,
  and Pressé]{doi:10.1063/1.5008842}
Sgouralis,~I.; Whitmore,~M.; Lapidus,~L.; Comstock,~M.~J.; Pressé,~S. Single
  molecule force spectroscopy at high data acquisition: A Bayesian
  nonparametric analysis. \emph{The Journal of Chemical Physics} \textbf{2018},
  \emph{148}, 123320\relax
\mciteBstWouldAddEndPuncttrue
\mciteSetBstMidEndSepPunct{\mcitedefaultmidpunct}
{\mcitedefaultendpunct}{\mcitedefaultseppunct}\relax
\EndOfBibitem
\bibitem[Kilic \latin{et~al.}(2021)Kilic, Sgouralis, and Pressé]{KILIC2021409}
Kilic,~Z.; Sgouralis,~I.; Pressé,~S. Generalizing HMMs to Continuous Time for
  Fast Kinetics: Hidden Markov Jump Processes. \emph{Biophysical Journal}
  \textbf{2021}, \emph{120}, 409--423\relax
\mciteBstWouldAddEndPuncttrue
\mciteSetBstMidEndSepPunct{\mcitedefaultmidpunct}
{\mcitedefaultendpunct}{\mcitedefaultseppunct}\relax
\EndOfBibitem
\bibitem[Tavakoli \latin{et~al.}(2020)Tavakoli, Jazani, Sgouralis, Shafraz,
  Sivasankar, Donaphon, Levitus, and Press\'e]{PhysRevX.10.011021}
Tavakoli,~M.; Jazani,~S.; Sgouralis,~I.; Shafraz,~O.~M.; Sivasankar,~S.;
  Donaphon,~B.; Levitus,~M.; Press\'e,~S. Pitching Single-Focus Confocal Data
  Analysis One Photon at a Time with Bayesian Nonparametrics. \emph{Phys. Rev.
  X} \textbf{2020}, \emph{10}, 011021\relax
\mciteBstWouldAddEndPuncttrue
\mciteSetBstMidEndSepPunct{\mcitedefaultmidpunct}
{\mcitedefaultendpunct}{\mcitedefaultseppunct}\relax
\EndOfBibitem
\bibitem[Kilic \latin{et~al.}(2021)Kilic, Sgouralis, Heo, Ishii, Tahara, and
  Pressé]{KILIC2021100409}
Kilic,~Z.; Sgouralis,~I.; Heo,~W.; Ishii,~K.; Tahara,~T.; Pressé,~S.
  Extraction of rapid kinetics from smFRET measurements using integrative
  detectors. \emph{Cell Reports Physical Science} \textbf{2021}, \emph{2},
  100409\relax
\mciteBstWouldAddEndPuncttrue
\mciteSetBstMidEndSepPunct{\mcitedefaultmidpunct}
{\mcitedefaultendpunct}{\mcitedefaultseppunct}\relax
\EndOfBibitem
\bibitem[Bishop(2006)]{bishop}
Bishop,~C.~M. \emph{Pattern Recognition and Machine Learning (Information
  Science and Statistics)}; Springer-Verlag: Berlin, Heidelberg, 2006\relax
\mciteBstWouldAddEndPuncttrue
\mciteSetBstMidEndSepPunct{\mcitedefaultmidpunct}
{\mcitedefaultendpunct}{\mcitedefaultseppunct}\relax
\EndOfBibitem
\bibitem[Sivia(2006)]{sivia}
Sivia,~D.~M. \emph{Data Analysis : A Bayesian tutorial}; Oxford University
  Press: Oxford, New York, 2006\relax
\mciteBstWouldAddEndPuncttrue
\mciteSetBstMidEndSepPunct{\mcitedefaultmidpunct}
{\mcitedefaultendpunct}{\mcitedefaultseppunct}\relax
\EndOfBibitem
\bibitem[van~de Schoot \latin{et~al.}(2021)van~de Schoot, Depaoli, King,
  Kramer, Märtens, Tadesse, Vannucci, Gelman, Veen, Willemsen, and
  Yau]{van_de_schoot_bayesian_2021}
van~de Schoot,~R.; Depaoli,~S.; King,~R.; Kramer,~B.; Märtens,~K.;
  Tadesse,~M.~G.; Vannucci,~M.; Gelman,~A.; Veen,~D.; Willemsen,~J.
  \latin{et~al.}  Bayesian statistics and modelling. \emph{Nature Reviews
  Methods Primers} \textbf{2021}, \emph{1}, 1\relax
\mciteBstWouldAddEndPuncttrue
\mciteSetBstMidEndSepPunct{\mcitedefaultmidpunct}
{\mcitedefaultendpunct}{\mcitedefaultseppunct}\relax
\EndOfBibitem
\bibitem[ATT()]{ATTO647N}
ATTO647N Dye Specifications.
  \url{https://www.atto-tec.com/fileadmin/user_upload/Katalog_Flyer_Support/ATTO_647N.pdf},
  Accessed: 2021-09-29\relax
\mciteBstWouldAddEndPuncttrue
\mciteSetBstMidEndSepPunct{\mcitedefaultmidpunct}
{\mcitedefaultendpunct}{\mcitedefaultseppunct}\relax
\EndOfBibitem
\bibitem[Gautier \latin{et~al.}(2001)Gautier, Tramier, Durieux, Coppey, Pansu,
  Nicolas, Kemnitz, and Coppey-Moisan]{GAUTIER20013000}
Gautier,~I.; Tramier,~M.; Durieux,~C.; Coppey,~J.; Pansu,~R.; Nicolas,~J.-C.;
  Kemnitz,~K.; Coppey-Moisan,~M. Homo-FRET Microscopy in Living Cells to
  Measure Monomer-Dimer Transition of GFP-Tagged Proteins. \emph{Biophysical
  Journal} \textbf{2001}, \emph{80}, 3000--3008\relax
\mciteBstWouldAddEndPuncttrue
\mciteSetBstMidEndSepPunct{\mcitedefaultmidpunct}
{\mcitedefaultendpunct}{\mcitedefaultseppunct}\relax
\EndOfBibitem
\bibitem[Liesche \latin{et~al.}(2015)Liesche, Grußmayer, Ludwig, Wörz, Rohr,
  Herten, Beaudouin, and Eils]{LIESCHE20152352}
Liesche,~C.; Grußmayer,~K.; Ludwig,~M.; Wörz,~S.; Rohr,~K.; Herten,~D.-P.;
  Beaudouin,~J.; Eils,~R. Automated Analysis of Single-Molecule Photobleaching
  Data by Statistical Modeling of Spot Populations. \emph{Biophysical Journal}
  \textbf{2015}, \emph{109}, 2352--2362\relax
\mciteBstWouldAddEndPuncttrue
\mciteSetBstMidEndSepPunct{\mcitedefaultmidpunct}
{\mcitedefaultendpunct}{\mcitedefaultseppunct}\relax
\EndOfBibitem
\bibitem[Scheible \latin{et~al.}(2015)Scheible, Ong, Woehrstein, Jungmann, Yin,
  and Simmel]{compact_DNA}
Scheible,~M.~B.; Ong,~L.~L.; Woehrstein,~J.~B.; Jungmann,~R.; Yin,~P.;
  Simmel,~F.~C. A Compact DNA Cube with Side Length 10 nm. \emph{Small}
  \textbf{2015}, \emph{11}, 5200--5205\relax
\mciteBstWouldAddEndPuncttrue
\mciteSetBstMidEndSepPunct{\mcitedefaultmidpunct}
{\mcitedefaultendpunct}{\mcitedefaultseppunct}\relax
\EndOfBibitem
\end{mcitethebibliography}


\end{document}


\maketitle
\section{S1. Verification of Sampler Using Synthetic Data}

The figure below shows a bivariate density plot for rates and excitation probabilities learned using the sampler described in the main manuscript for synthetically generated data. The six rates that were used are 0.01, 0.1, 0.5, 2.0, 4.0, and 10.0 ns$^{-1}$, and the corresponding excitation probabilities are 0.2, 0.3, 0.1, 0.2, 0.1, and 0.1. These ground truth values are shown as orange dots in the plot. Our Markov-Chain Monte-Carlo sampler accurately predicts most of these rates and probabilities, however, rates that are not well separated in magnitude have much larger spread in their distributions. For instance, the learned distribution around the rates corresponding to 2.0 and 4.0 ns$^{-1}$ is much wider in the plot below.

\begin{figure}[h!]
    \centering
    \includegraphics[width =0.7 \textwidth]{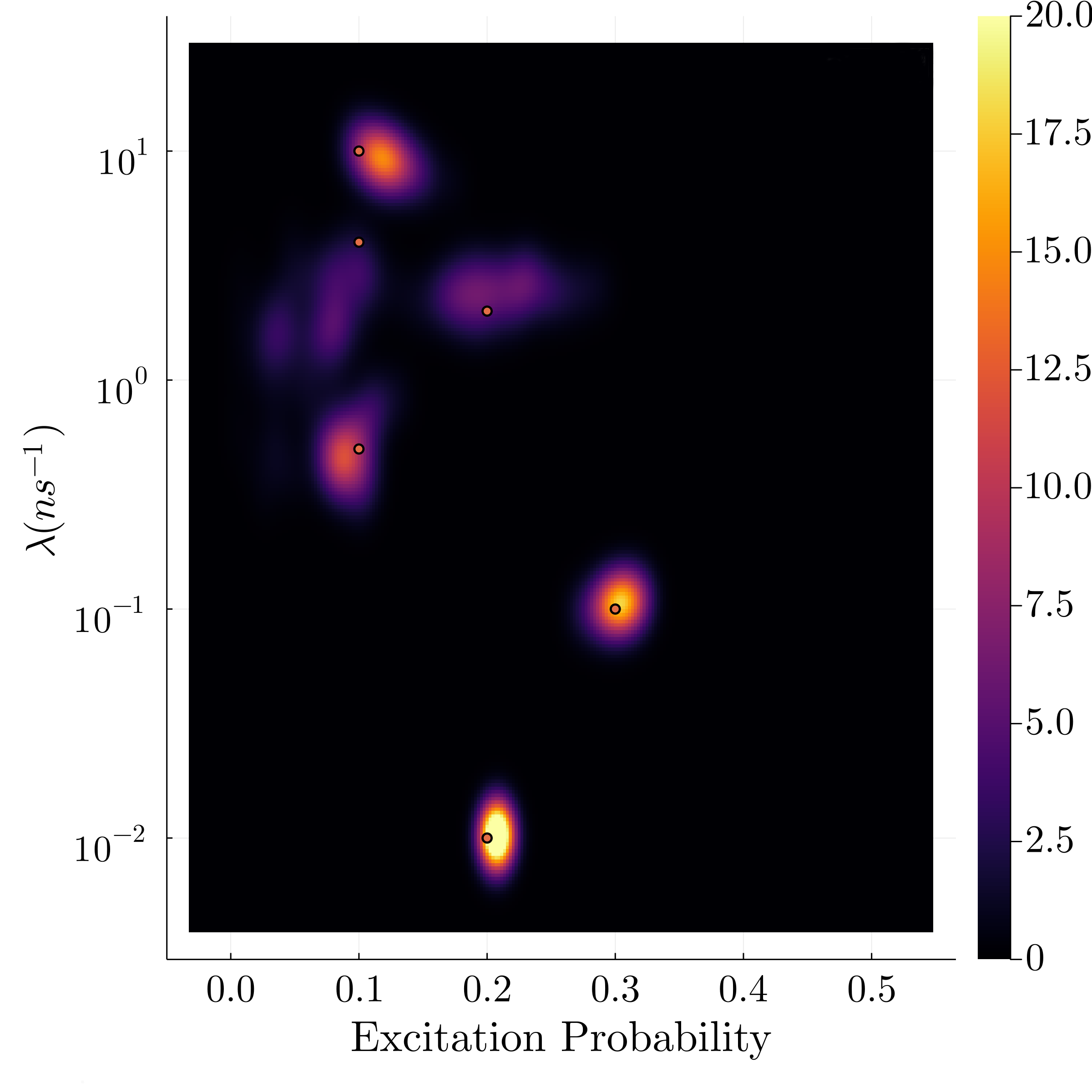}
	\caption{\textbf{Robustness Test.} Bivariate density plot for a system 
	with six photophysical states with escape rates 0.01, 0.1, 0.5, 2.0, 
	4.0, and 10.0 ns$^{-1}$, and the corresponding excitation probabilities 
	are 0.2, 0.3, 0.1, 0.2, 0.1, and 0.1. The ground truth is shown as 
	dots. }
    \label{ }
\end{figure}

\section{S2. Results Using Metropolis-Hastings Algorithm}

\begin{figure*}[h!]
	\includegraphics[width=0.8\textwidth]{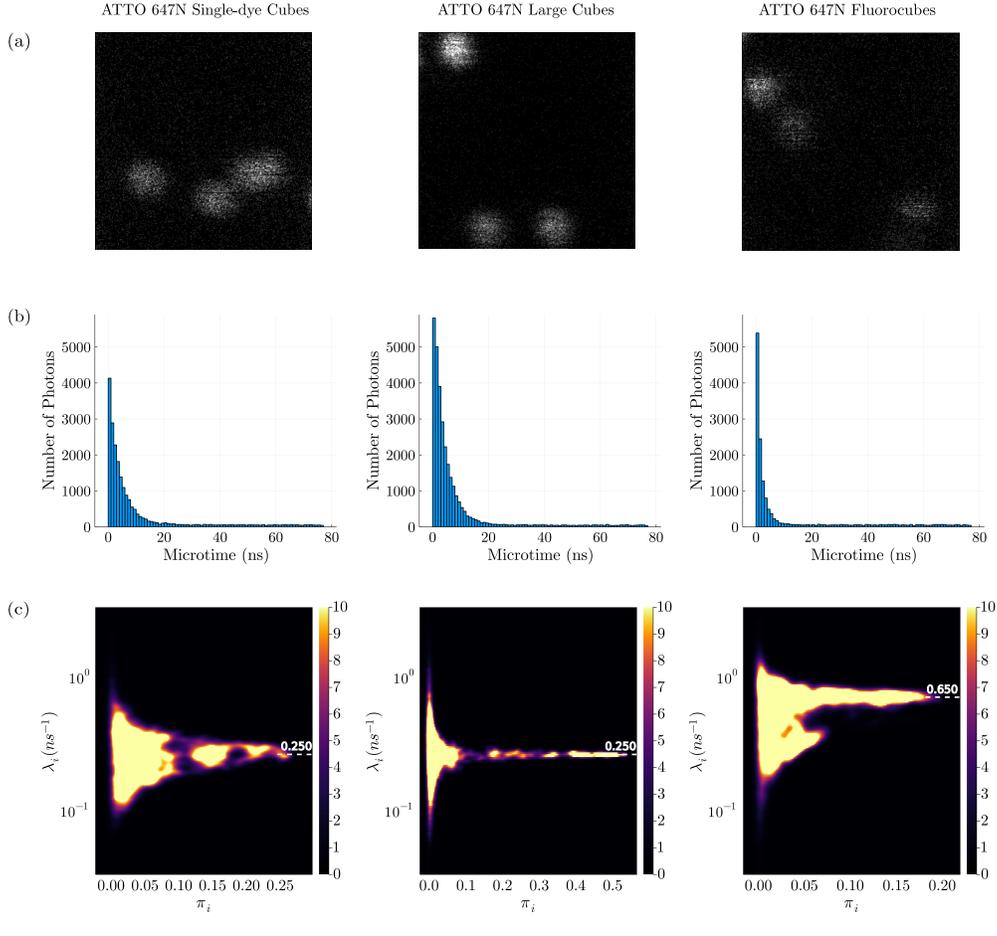}]
	\caption{\textbf{Experimental Data Analysis Using Metropolis-Hastings.} 
	In row (a), we have three $256 \times 256$ pixels raster-scanned images 
	of samples containing three cubes labeled with ATTO 647N dyes, each. 
	The images on the left and the right are for single-dye cubes and 
	fluorocubes illuminated with 80\% maximum laser power, respectively. 
	However, the center image for large-cubes was produced using only 30\% 
	laser power. The laser power is varied in order to obtain approximately 
	equal photon numbers per probe per  experimental run. Histograms for 
	photon arrival times (microtimes) recorded from these images are shown 
	in row (b). In row (c), each panel shows a bivariate posterior for 
	escape rates $\lambda_i$ (log scale) and corresponding probabilities 
	$\pi_i$ as in Eq.~5 of the main manuscript. We smooth out the bivariate 
	plots using the kernel density estimation (KDE) tool available in 
	Julia. The left panel shows the distribution for single-dye cubes. The 
	distribution is concentrated about $\lambda_i \approx 0.25$ ns$^{-1}$ 
	or a lifetime of around $1/\lambda_i$ $= 4.0$ ns. $\pi_b \approx 0.10$ 
	and $\pi_0 \approx 0.49$ in this case. The plot for large cubes in the 
	center panel shows no change in lifetime, suggesting insignificant 
	interactions. $\pi_b \approx 0.08$ and $\pi_0 \approx 0.37$ for these 
	cubes. In the rightmost panel for fluorocubes, the peak moves towards a 
	significantly shorter lifetime of around 1.54 ns. The background 
	probability $\pi_b$ here is 0.12, slightly higher than that of 
	non-interacting cubes, and $\pi_0 \approx 0.59$.  }
	\label{distributions}
\end{figure*}

\newpage
\section{S3. Results For Cy3 Fluorocubes}
\begin{figure*}[h!]
\includegraphics[width=0.65\textwidth]{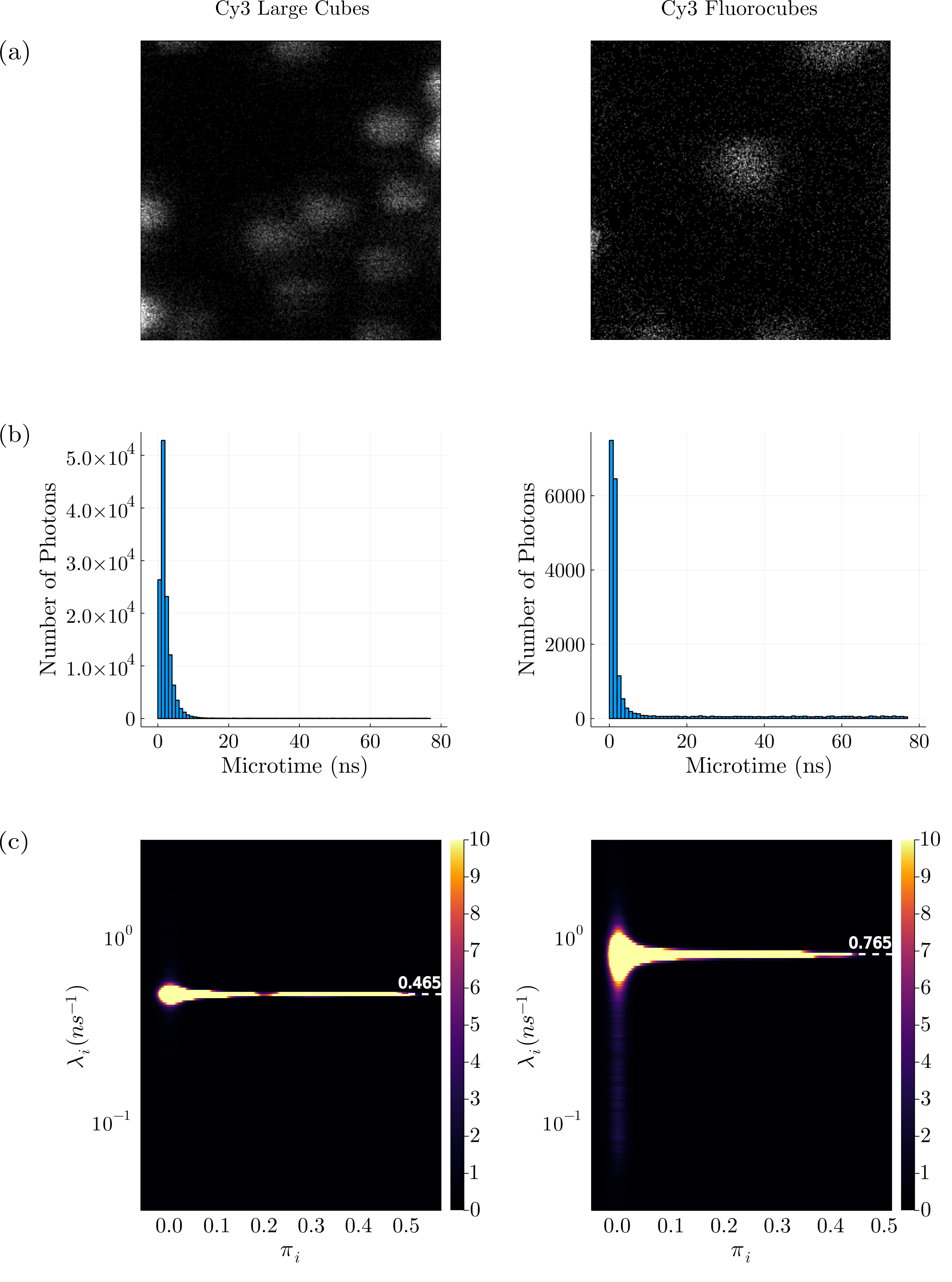}
	\caption{\textbf{Experimental Data Analysis.} Here, we show raster 
	scanned images (a), microtime histograms (b), and learned posterior 
	distributions (c) for probes labeled with Cy3 dyes. Each panel in row 
	(c) shows a bivariate posterior for escape rates $\lambda_i$ (log 
	scale) and corresponding probabilities, learned from a single 
	experimental run with  samples containing multiple fluorescent probes 
	of the same type (fluorocubes or large cubes) in the field of view. In 
	the left panel, we show the distribution for large cubes and 
	illuminated with 85\% of the maximum laser power. The distribution is 
	concentrated about $\lambda_i \approx 0.465$ ns$^{-1}$ or a lifetime of 
	around $1/\lambda_i$ $= 2.15$ ns.  $\pi_b \approx 0.03$ and $\pi_0 
	\approx 0.09$ in this case.  In the second panel on the right, for 
	fluorocubes labeled with six dyes illuminated with 85\% laser power, 
	the peak moves towards a significantly shorter lifetime of around 1.31 
	ns. The background probability is about 0.15 here and 27 \% of the 
	pulses do not lead to any photon detections.
}
	\label{distributions}
\end{figure*}

\newpage
\section{S4. Experimental Details}
\begin{table}
\scalebox{0.6}{
\begin{tabular}{ |c|c|c|}
 \hline
 \textbf{Name} & \textbf{Sequence and modification} & Vendor \\
 \hline
 FC\_SC\_01\_Cy3 &  /5Cy3/ATGAGGTGTATGTGTAGAGTGATGGATGTAGT/3Cy3/ & IDT \\
\hline
FC\_SC\_02\_Cy3 & /5Cy3/AGGATGAGTGAGAGTGAGATGAGAGTAGATGT/3Cy3/
 & IDT \\
 
\hline
FC\_St\_02\_Cy3 & /5Cy3/CACTCTCACACCTCATACATCTACCATCACTC/3Cy3/ & IDT \\

\hline
FC\_SC\_01\_ATTO647N & /5ATTO647NN/ATGAGGTGTATGTGTAGAGTGATGGATGTAGT/3ATTO647NN/
& IDT \\

\hline
FC\_SC\_02\_ATTO647N & /5ATTO647NN/AGGATGAGTGAGAGTGAGATGAGAGTAGATGT/3ATTO647NN/ & IDT \\
\hline
FC\_St\_02\_ATTO647N & /5ATTO647NN/CACTCTCACACCTCATACATCTACCATCACTC/3ATTO647NN/ & IDT \\
\hline 
SD\_SC\_01\_UN & TTATGAGGTGTATGTGTAGAGTGATGGATGTAGTTT & IDT \\
\hline
SD\_SC\_02\_UN & TTAGGATGAGTGAGAGTGAGATGAGAGTAGATGTTT & IDT \\
\hline
SD\_St\_02\_ATT0647N & /5ATTO647NN/CACTCTCACACCTCATACATCTACCATCACTCTT & IDT \\
\hline
SD\_St\_02\_Cy3 & /5Cy3/CACTCTCACACCTCATACATCTACCATCACTCTT & IDT \\
\hline
FC\_St\_01\_Bio & Biotin TACACATACTCATCCTACTACATCTCTCATCTTT & IDT \\
\hline
LC\_SC\_01 & TTGAAAATTATCTCGATAAGCAGAAGGACCTGTATAACTGGCAAGAGACAAGGCCGCTTCAGAA & IDT \\
\hline
LC\_SC\_02 & AGGATAGCCGGACCGTATTAATGCCGCGCCAACGGTTTCCCGGACCTAGTGTCTATCAAGTCTA & IDT \\
\hline
LC\_SC\_03 & TTCTATGAAACCATTCTCGGGTCGAGCGGGTCACTGTTGTGACCTACGAGAAGCGTATAGATGT & IDT \\
\hline 
LC\_SC\_04 & TCCGCGCGAATAGCTCACAGGCGAACTACGTATGAATTGGTTTAAACGCTCCTCGGGAATTAAT & IDT \\
\hline
LC\_SC\_05 & ACGACAGGTGGCAAACCACCTCCGATGTCAGCGCCGCATACCCATTCACTGTGAATTTCCACAC & IDT \\
\hline
LC\_SC\_06 & CGAGGATTCGCAGGTCCATGGGATTCACCAAGCTCGTATACACCCTGATTCTCCATGGCAGCGC & IDT \\
\hline
LC\_SC\_07 & GTAAGTTGAAGTAGGAAGCTTTTTCTAGCCATAGCATCGACACTACGACCTGCTTTTCGACACA & IDT \\
\hline
LC\_SC\_08 & GGACTGCATTCTGGACAGTAACTGCATTAACTACGTGCTCCCAACATAAGTGACGTCCTCAGCA & IDT \\
\hline
LC\_St\_01 & TTTGCTGAGGTGGAAATTTT & IDT \\
\hline
LC\_St\_02 & TTCCGGGAAACCGTTGGCCCTTCTGCTCGCCTGTCGTAGGTCGGGT & IDT \\
\hline
LC\_St\_03\_ATTO647N & GGTTCTGCGAATCCTCGGTGACGTCACTTCCTACTTCAACTTACTT/ATTO647N & IDT \\
\hline
LC\_St\_03\_Cy3 & GGTTCTGCGAATCCTCGGTGACGTCACTTCCTACTTCAACTTACTT/Cy3 & IDT \\
\hline
LC\_St\_04 & GTATGCGGCGCTCAGTTACTTCGTAGTGTCGATGCTTT & IDT \\
\hline
LC\_St\_05\_Bio & /5Biosg/TTCCATGGACTCATAGAATT & IDT \\
\hline
LC\_St\_06 & TTGAGAATCAACAACAGTTT & IDT \\
\hline
LC\_St\_07\_ATTO647N & TTACATCTATCACTAGGTTT/ATTO647N & IDT \\
\hline
LC\_St\_07\_Cy3 & TTACATCTATCACTAGGTTT/Cy3 & IDT \\
\hline
LC\_St\_08 & TTTACGTAGTTTATCGAGTT & IDT \\
\hline
LC\_St\_09\_ATTO647N & ATTO647N/TTCCGGCTATCCTTTCTGAAGCGGCCCCGAGGAGGAAT & IDT \\
\hline
LC\_St\_09\_Cy3 & Cy3/TTCCGGCTATCCTTTCTGAAGCGGCCCCGAGGAGGAAT & IDT \\
\hline
LC\_St\_10 & GTATACGAGCTTAAAAAGCTTATGTTGGGAGCACGTTT & IDT \\
\hline
LC\_St\_11 & TTTGCCAGTTATACAGGTGCGGCATTCGACCCGACGTTTAAAATGG & IDT \\
\hline
LC\_St\_12 & TTAGGTGGTTCGCGCGGATT & IDT \\
\hline
LC\_St\_13 & TTAGTTAATGGACATCGGTT & IDT \\
\hline
LC\_St\_14\_ATTO647N & ATTO647N/TTATAATTTTCAATAGACTTGATAGAACGCTTCTGAGCTATTTGCC & IDT \\
\hline
LC\_St\_14\_Cy3 & Cy3/TTATAATTTTCAATAGACTTGATAGAACGCTTCTGAGCTATTTGCC & IDT \\
\hline
LC\_St\_15 & TTGACCCGCTAATACGGTTT & IDT \\
\hline
LC\_St\_16\_ATTO647N & ACCTGTCGTGCGAAAGCAGGGTCCAGAATGCAGTCCTT/ATTO647N & IDT \\
\hline
LC\_St\_16\_Cy3 & ACCTGTCGTGCGAAAGCAGGGTCCAGAATGCAGTCCTT/Cy3 & IDT \\
\hline
LC\_St\_17 & TTATGGCTAGGGTGAATCTT & IDT \\
\hline
LC\_St\_18 & TTCACAGTGACCAATTCATT & IDT \\
\hline
LC\_St\_19 & TTTGTGTCGACTGCCATGTT & IDT \\
\hline
LC\_St\_20\_ATTO647N & TTATTAATTCTTGTCTCTTT/ATTO647N & IDT \\
\hline
LC\_St\_20\_Cy3 & TTATTAATTCTTGTCTCTTT/Cy3 & IDT \\
\hline
\end{tabular}}
\caption{Sequences for Assembly of DNA FluoroCubes and Large Cubes.}
\end{table}